\documentclass[fleqn,usenatbib]{mnras}

\usepackage{newtxtext,newtxmath}

\usepackage[T1]{fontenc}
\usepackage{xcolor}
\usepackage{lscape}

\DeclareRobustCommand{\VAN}[3]{#2}
\let\VANthebibliography\thebibliography
\def\thebibliography{\DeclareRobustCommand{\VAN}[3]{##3}\VANthebibliography}

\usepackage{graphicx}	
\usepackage{amsmath}	

\newcommand{\Ha}{H$\alpha$}
\newcommand{\Lya}{Ly$\alpha$}
\newcommand{\HI}{H\,\textsc{i}}
\newcommand{\HII}{H\,\textsc{ii}}
\newcommand{\HeI}{He\,\textsc{i}}

\newcommand{\CI}{C\,\textsc{i}}
\newcommand{\CaII}{Ca\,\textsc{ii}}

\title[Featureless stars]{Featureless stars: Flux Calibration for Extremely Large Telescopes}

\author[Cooke et al.]{
Ryan J. Cooke,$^{1,2}$\thanks{E-mail: ryan.j.cooke@durham.ac.uk (RJC)}
Nao Suzuki,$^{3,4,5}$
J. Xavier Prochaska$^{5,6,7,8}$
\\
$^{1}$Centre for Extragalactic Astronomy, Durham University, South Road, Durham DH1 3LE, UK\\
$^{2}$Department of Physics, Durham University, South Road, Durham DH1 3LE, UK\\
$^{3}$Physics Department, Florida State University,
77 Chieftan Way, Tallahassee, FL 32306, USA\\
$^{4}$Physics Division, Lawrence Berkeley National Laboratory,
1 Cyclotron Road, Berkeley, CA 94720, USA\\
$^{5}$Kavli Institute for the Physics and Mathematics of the Universe, University of Tokyo, Kashiwa 277-8583, Japan\\
$^{6}$Department of Astronomy and Astrophysics, University of California at Santa Cruz, 1156 High Street, Santa Cruz, CA 95064, USA\\
$^{7}$University of California Observatories, Lick Observatory, 1156 High Street, Santa Cruz, CA 95064, USA\\
$^{8}$Division of Science, National Astronomical Observatory of Japan,2-21-1 Osawa, Mitaka, Tokyo 181-8588, Japan\\
}

\date{Accepted XXX. Received YYY; in original form ZZZ}

\pubyear{2024}

\begin{document}
\label{firstpage}
\pagerange{\pageref{firstpage}--\pageref{lastpage}}
\maketitle

\begin{abstract}
The spectrophotometric flux calibration of recent spectroscopic surveys has reached a limiting systematic precision of approximately $1-3$ percent, and is often biased near the wavelengths associated with \HI\ Balmer absorption. As we prepare for the next generation of imaging and spectroscopic surveys, and high-precision cosmology experiments, we must find a way to address this systematic. Towards this goal, we have identified a global network of 29 bright ($G<17.5$) featureless white dwarf stars that have a spectral energy distribution consistent with an almost pure blackbody form over the entire optical and near-infrared wavelength range. Based on this sample, we have computed the systematic uncertainty and AB magnitude offsets associated with Gaia, SDSS, SMSS, PanSTARRS, DES, and 2MASS, and we have also checked the consistency of our objects with both GALEX and WISE. The magnitude range of the featureless stars reported here are ideally suited to observations taken with the forthcoming generation of extremely large telescopes, as well as calibrating the survey data acquired by the Rubin, Euclid and Roman observatories. Finally, all of the high-precision spectrophotometric standard stars reported here have been included in the latest release of the \textsc{PypeIt} data reduction pipeline.
\end{abstract}

\begin{keywords}
cosmological parameters -- cosmology: observations -- standards -- white dwarfs -- catalogues
\end{keywords}


\section{Introduction}


Determining the absolute brightness of an astronomical object is one of the oldest disciplines in observational astronomy, and is one of the cornerstones of modern astronomical surveys. Photometric calibration is achieved by mapping fluxes from detector units to physical units. This ultimately requires a stable calibration source (i.e. non-variable) over the waveband of interest. When performing a spectrophotometric calibration, one also requires knowledge of the intrinsic spectral shape of the source and its stability. The most common approach is to use observations of a well-characterised star, that is often modelled with a theoretical understanding of the stellar atmosphere. Perhaps the most well-characterised stars are white dwarfs, which can reach a photometric precision of $\sim1$ percent at optical wavelengths \citep{Bohlin2014,Bohlin2020,Axelrod2023,Bohlin2025}. Nearby unextincted G-type stars could provide a comparable level of precision ($\sim1$ percent; \citealt{Rieke2024}), while A-type stars can reach a precision of a few percent, but often exhibit variability \citep{AllendePrieto2016}. Alternatively, laboratory calibrated sources, such as emissive reference spheres \citep{Price2004}, offer an opportunity to provide an absolute flux calibration to the level of $\sim1$ percent.

For most astronomical applications, the precision delivered by well-calibrated white dwarf stars is more than sufficient. The most widely used flux standard stars are based on the \emph{Hubble Space Telescope} (\emph{HST}) CALSPEC database \citep{Bohlin2025}.\footnote{The latest CALSPEC files can be accessed from the following website: \\ \url{https://www.stsci.edu/hst/instrumentation/reference-data-for-calibration-and-tools/astronomical-catalogs/calspec}} This compilation of flux standard stars is estimated to be accurate to the level of $\sim1$ percent, and has been widely adopted by the most recent generation of wide-area imaging and spectroscopic surveys. The longevity of all survey data are limited by the accuracy of the calibration. For example, some of the most recent imaging surveys, including Pan-STARRS \citep{Chambers2016,Magnier2020} and the Dark Energy Survey (DES; \citealt{DES}), report a flux calibration precision of $\sim1$ percent \citep{Chambers2016,Burke2018}. The situation is somewhat more critical for spectroscopic surveys. The Sloan Digital Sky Survey (SDSS) spectroscopic survey reports a spectrophotometric calibration of $1-2$ percent over most of the wavelength coverage, with systematic features at the level of $2-3$ percent appearing near wavelengths corresponding to the \HI\ Balmer lines \citep{Lan2018}, attributed to spectral features in the calibration standards. We also note that the SDSS flux calibration appears to deviate by $5-10\%$ at wavelengths $<3700$\,\AA\ \citep{Kamble2020}.

There is a growing need to identify and model flux standard stars with a precision that breaks the $\sim1$ percent barrier \citep{Kent2009}. Photometric calibration is one of the dominant sources of uncertainty in supernova samples \citep{Conley2011,Scolnic2014} and limits the accuracy of tracing the expansion history of the Universe. Other cosmological probes are limited by spectrophotometric precision, such as the determination of photometric redshifts to trace the growth of structure \citep{Schmidt2020} and measures of the primordial helium-4 abundance from metal-poor \HII\ regions \citep{OliveSkillman2004,Izotov2007}. Flux calibration is particularly difficult near strong absorption features that are present in the spectrum of the calibration star, and this can introduce biases in science targets at these wavelengths, particularly affecting the spectra of objects in the local Universe (i.e. the Milky Way and the Local Group).

The current gold standard of optical flux calibration is based on a non local thermodynamic equilibrium model fit to three white dwarf stars (G191$-$B2B, GD\,153, and GD\,71; \citealt{Bohlin2020}) that have been scaled to the absolute laboratory calibrated flux of Vega at a vacuum wavelength of 5557.5\,\AA\ in the optical \citep{Megessier1995} and with the Midcourse Space Experiment (MSX) calibration of Sirius in the mid-infrared \citep{Price2004}. The absolute calibration accuracy is $\sim1$ percent, while the relative accuracy between any two wavelengths is somewhat better, but can be as large as 1 percent depending on the two wavelengths being compared. High quality standard stars in the near-infrared are greatly needed, and this is one of the key goals of the James Webb Space Telescope absolute flux calibration programme \citep{Gordon2022}. Furthermore, the future generation of extremely large telescopes \citep{PadovaniCirasuolo2023} will require somewhat fainter flux standard stars than those currently used, particularly for near-infrared adaptive optics imaging observations \citep{Boyd2024,Bohlin2025,GentileFusillo2025}. Furthermore, under normal operating modes, most historic standard stars ($m < 15$) will saturate the detectors of forthcoming surveys on large aperture telescopes, including the Legacy Survey of Space and Time (LSST), Euclid, and the Nancy Grace Roman Telescope.

In order to break the one percent photometric calibration barrier while using standard stars that are suitable for both current and future observatories, we need to identify a network of stars that are independent of the white dwarf models that are used to calibrate the stars from the ultraviolet to the near-infrared. Towards this goal, \citet{SuzukiFukugita2018} identified a small number of white dwarf stars that appear to be featureless over the optical wavelength range, and obey a near perfect blackbody distribution from the far-ultraviolet to the mid-infrared. The first blackbody star to be identified was originally targeted as a quasar candidate by the SDSS spectroscopic survey \citep{Ross2012}, and subsequently identified as a white dwarf star \citep{Paris2014}. This serendipitous discovery prompted \citet{SuzukiFukugita2018} to examine 798,593 spectra to identify just 17 featureless blackbody stars. These stars are primarily DC-type white dwarf stars with effective temperatures $\lesssim11,000~{\rm K}$ \citep{SerenelliRohrmannFukugita2019}, therefore representing the cooler analogues of DB white dwarf stars.

Featureless stars represent a promising calibrator that will allow for precise (i.e. relative) and accurate (i.e. absolute) flux calibration for future extremely large telescope facilities, astronomical survey missions, and dedicated cosmology experiments. However, the present sample of blackbody stars is limited to northern declinations ($\delta\gtrsim-1^{\circ}$), faint magnitudes ($m_{\rm G}>17.1$), and are relatively rare. In this paper, we develop a strategy to efficiently identify and characterise the best and brightest featureless stars that can be used for spectrophotometric flux calibration.

This paper is outlined as follows: In Section~\ref{sec:candidates}, we outline our candidate selection strategy, based on \textit{Gaia} photometry, astrometry, and spectroscopy. We describe our observational programme in Section~\ref{sec:obs}, including the data reduction recipes used. In Section~\ref{sec:analysis} we describe our equivalent width analysis, and our final set of spectroscopic criteria that must be satisfied to classify a white dwarf as featureless for the purpose of flux calibration. In Section~\ref{sec:blackbodyparams} we cross-match our featureless star sample with wide-area sky surveys, and perform parametric fits to the available photometry of our stars. We also provide estimates of the zeropoint offsets of each survey and filter that are needed to bring the catalogue magnitudes onto the AB magnitude scale, together with the intrinsic uncertainty of each filter magnitude. Finally, we also revisit the best selection box to identify new featureless stars in the future, before drawing our main conclusions in Section~\ref{sec:conc}.

\section{candidate selection}
\label{sec:candidates}

\begin{figure}
	\includegraphics[width=\columnwidth]{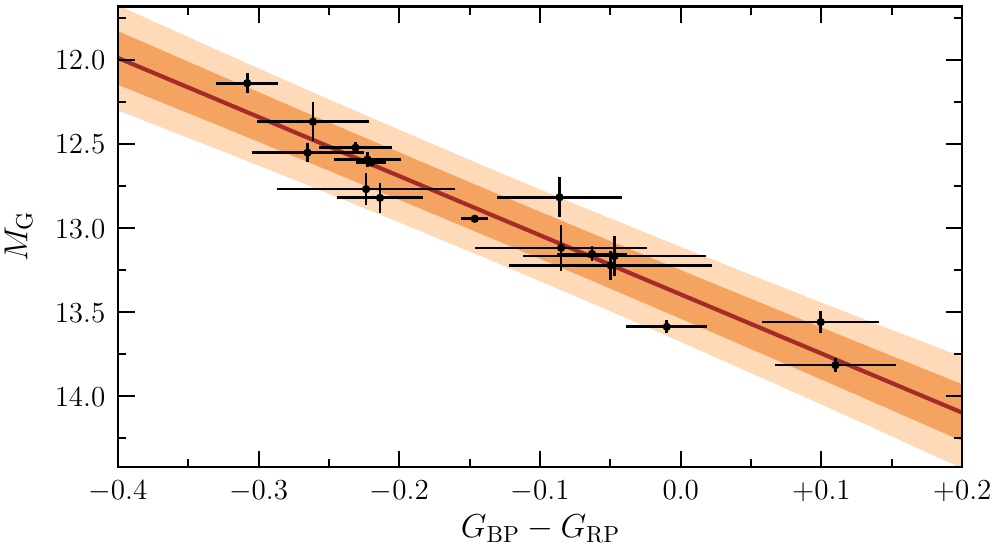}
    \caption{The \citet{SuzukiFukugita2018} blackbody star sample (black points with error bars) occupy a tight relationship on the Gaia colour-magnitude diagram. The solid line represents a linear fit to the data, where the dark and light shaded regions indicate the 68 and 95 percent confidence intervals, respectively.}
    \label{fig:colmagSF18}
\end{figure}

Based on the \citet{SuzukiFukugita2018} sample of confirmed blackbody stars, together with \emph{Gaia} parallax and colour information (\citealt{Gaia2023}, see also \citealt{Hollands2018,OBrien2024}), we have identified that blackbody stars occupy an almost linear relationship on the colour-magnitude diagram\footnote{We use the AB zeropoints \citep{Riello2021}.} of the \emph{Gaia} experiment
(see Figure~\ref{fig:colmagSF18}). To determine the functional form of this relationship, we use the \texttt{emcee} software \citep{emcee} to perform a Monte Carlo Markov Chain (MCMC) fit to the absolute magnitudes of the \citet{SuzukiFukugita2018} blackbody stars given their \emph{Gaia} colours. We use 50 walkers, each with 5000 chains, and a burn-in of 500 chains. We calculate the autocorrelation of the samples to confirm that the parameters have converged to their best-fit values. We adopt a linear functional form with an intrinsic scatter in the absolute magnitude (see e.g. \citealt{Cooke2018}). The resulting best-fit parameters and their associated $68\%$ confidence intervals are:
\begin{eqnarray}
    \label{eqn:gaiafit}
    \!\!\!\!M_{\rm G} \!\!\!\!&=&\!\!\!\! (13.400\pm0.050) + (3.52\pm0.27)\cdot(G_{\rm BP} - G_{\rm RP})\\
    \!\!\!\!\sigma_{\rm int, G} \!\!\!\!&=&\!\!\!\! 0.118^{+0.036}_{-0.027}
\end{eqnarray}
where $M_{\rm G}$ is the absolute \emph{Gaia} $G$ band magnitude of the \citet{SuzukiFukugita2018} stars and $\sigma_{\rm int, G}$ is the intrinsic absolute magnitude scatter of the blackbody sample, i.e.\ we find
the observed scatter exceeds the observational error. 

We use this definition to define a boundary in the \emph{Gaia} colour-magnitude diagram that likely includes all nearby blackbody stars across the entire sky. The absolute magnitude boundary that we select represents the $95\%$ confidence interval of the relation given by Equation~\ref{eqn:gaiafit}, corresponding to roughly $\pm0.27~{\rm dex}$ above and below the mean relation. For completeness, we selected a colour boundary that includes all possible white dwarf stars, given by the range $-0.85 \leq G_{\rm BP}-G_{\rm RP} \leq +0.65$. Even though the \citet{SuzukiFukugita2018} blackbody sample only covers $G_{\rm BP}-G_{\rm RP}\gtrsim-0.33$, we have decided to extend this boundary to bluer colours to include the featureless star J1218+4148 ($G_{\rm BP}-G_{\rm RP}\simeq-0.67)$, which was not considered by these authors due to the lack of a turnover at FUV wavelengths.

Upon visual inspection of the \emph{Gaia} BP/RP spectra (see Section~\ref{sec:gaiabprpspec}) we decided to further extend our selection box to higher absolute magnitude when $-0.85 \leq G_{\rm BP}-G_{\rm RP} \leq -0.20$; this extra boundary includes additional featureless star candidates that appeared near the boundary of our original selection box. In the left panel of Figure~\ref{fig:gaiawd}, we show a two-dimensional histogram of all white dwarf stars within 220\,pc of the Sun, based on the Early Data Release 3 (EDR3)\footnote{Our candidate selection was done prior to the DR3 release, all other aspects of our analysis were performed on the \emph{Gaia} DR3 release.} \emph{Gaia} photometry and parallax measurements. We also plot our selection box that we use to identify candidate featureless white dwarf stars. In this paper, we are particularly interested in identifying the brightest featureless stars over the full sky. We therefore limit the apparent magnitude of our selection to all white dwarfs with $m_{\rm G}<17.5$. All of the \citet{SuzukiFukugita2018} blackbody stars have $m_{\rm G}>17.1$, and only two of their sample have $m_{\rm G}<17.5$. The magnitude range adopted by our new survey covers the brightest stars reported by \citet{SuzukiFukugita2018}, and is the ideal magnitude range for the forthcoming generation of extremely large telescopes ($16 < G < 18$). A total of 5102 candidate featureless stars meet the above criteria, and proceed to the next stage of classification.

\subsection{Gaia BP/RP spectroscopy}
\label{sec:gaiabprpspec}

\begin{figure*}
	\includegraphics[height=\columnwidth]{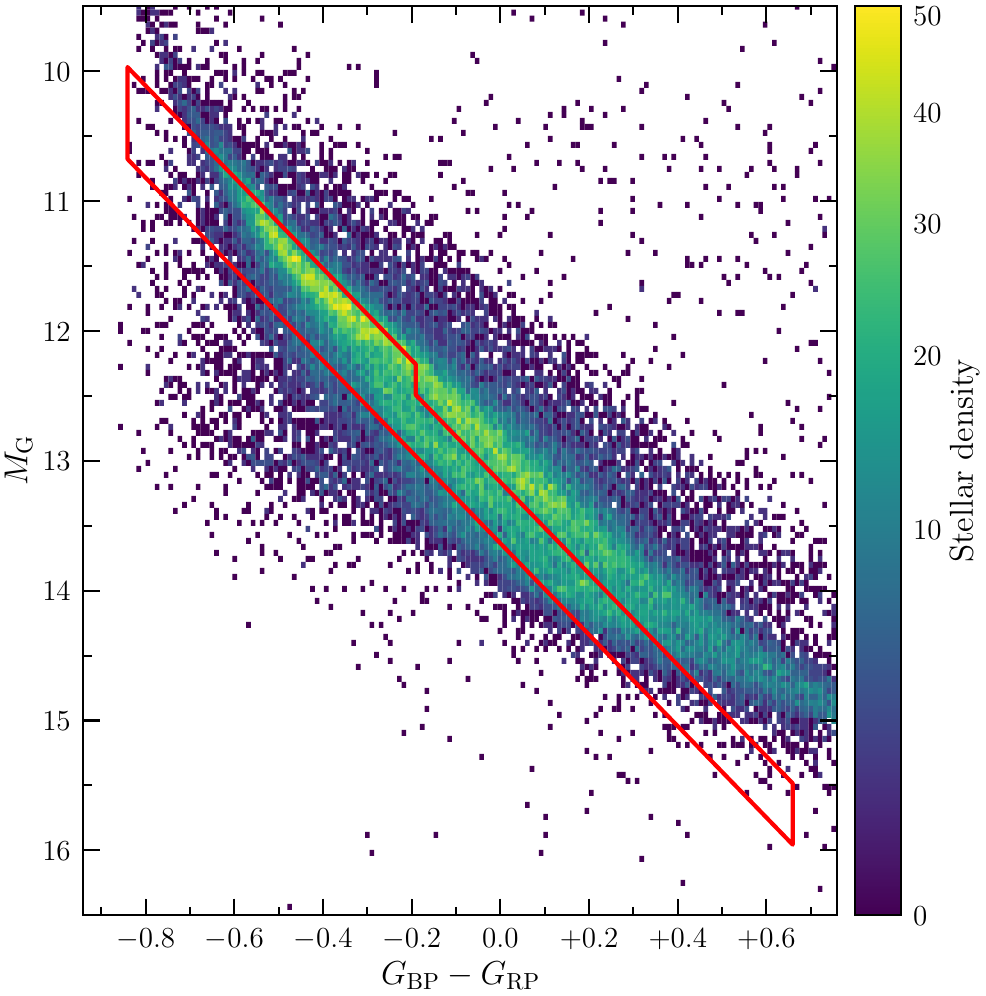}
        \hspace{0.05\columnwidth}
	\includegraphics[height=\columnwidth]{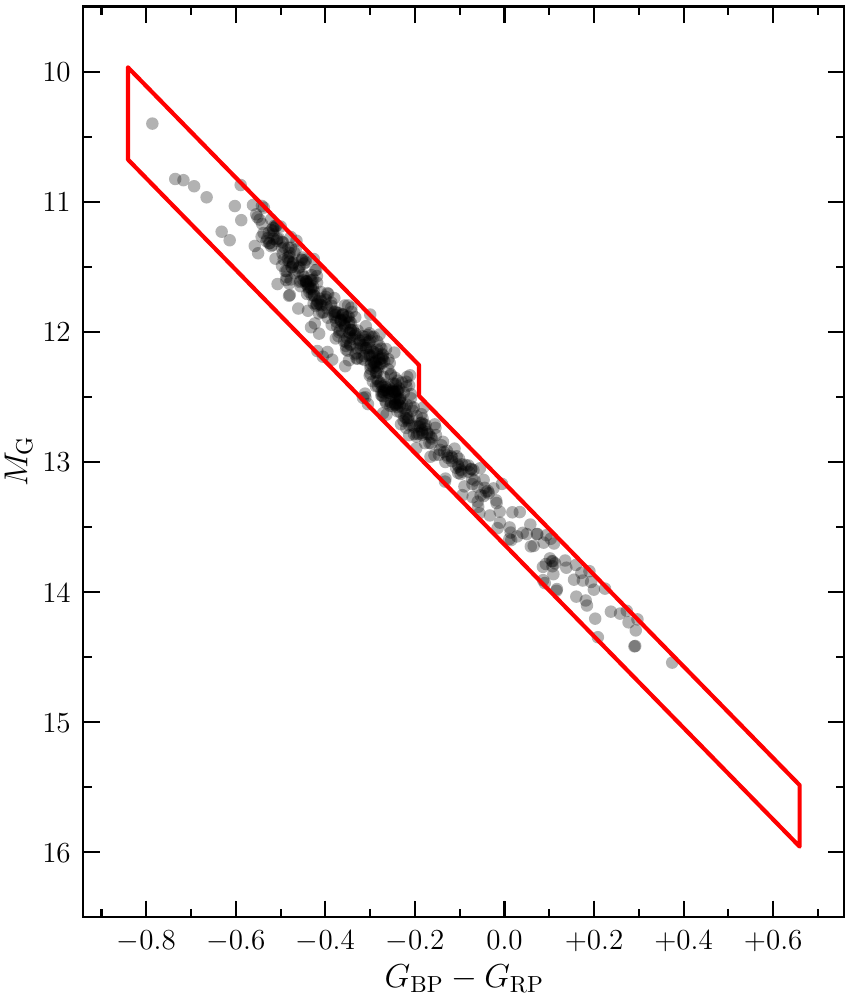}
    \caption{Both panels show the \emph{Gaia} colour-magnitude diagram for white dwarfs, with our candidate selection box shown as the red polygon. The left panel displays a histogram of the \emph{Gaia} white dwarfs within 220\,pc of the Sun, colour-coded by the stellar density. The right panel shows the distribution of our candidate featureless stars (black symbols) after removing white dwarf stars that display absorption lines, based on \emph{Gaia} BP/RP spectroscopy.}
    \label{fig:gaiawd}
\end{figure*}

Our selected magnitude limit ensures that there is a low resolution \emph{Gaia} BP/RP spectrum available to inspect if any candidates display obvious evidence of \HI\ Balmer and \HeI\ absorption. We visually inspected the \emph{Gaia} BP/RP spectroscopy of all 5102 candidate featureless white dwarfs. We removed any stars that exhibited clear and obvious spectral features due to either Balmer absorption lines or \HeI\ absorption lines.  Note that our pruning is conservative; we did not remove any candidates unless they clearly exhibit absorption features. This reduces our sample to 513 featureless candidates over the entire sky. We further pruned our sample, by removing:
(1) candidates with SDSS spectroscopy already available (21 candidates);
(2) candidates where SDSS or DSS imaging observations revealed a visually crowded region (40 candidates);
(3) the source is within 10 degrees of the Galactic plane (113 candidates);
(4) candidates with a declination $\delta>+72^{\circ}$, since most of our Northern objects were observed with the Shane/KAST spectrograph, which is not capable of observing at high declination (13 candidates); and
(5) candidates that are not detected by the Wide-Field Infrared Survey Explorer (WISE) imaging survey (40 candidates).
This resulted in a final sample of 286 candidates that we observed with telescopes in both the northern hemisphere (93 candidates with $\delta>0^{\circ}$) and southern hemisphere (193 candidates with $\delta<0^{\circ}$).

Our final sample of candidates are shown on a colour-magnitude diagram in the right panel of Figure~\ref{fig:gaiawd}. We note that during our first iteration of candidate selection, we noticed some candidates were grouped near the upper edge of the selection box at blue colours ($G_{\rm BP}-G_{\rm RP}<-0.2$). We therefore decided to extend our selection box, and repeat the candidate pruning, as described above. Note that the candidate numbers reported above are based on the entire selection box.
    
\section{Observations and Data Reduction}
\label{sec:obs}

In order to realise a full sky network of bright featureless stars, we observed our 286 candidate featureless stars with:
(i) the Kast spectrograph on the 3\,m Shane telescope at Lick observatory;
(ii) the Goodman spectrograph on the 4\,m Southern Astrophysical Research (SOAR) telescope \citep{Clemens2004};
(iii) the DEep Imaging Multi-Object Spectrograph (DEIMOS) at the W.~M.~Keck observatory; and
(iv) the Very Large Telescope (VLT) X-shooter spectrograph \citep{Vernet2011}. We complemented our observing programme with SDSS spectra of 21 candidates. After several observing runs, we further improved our strategy to select the highest priority featureless stars. In total, we obtained 367 spectra (including SDSS data, and observations of the same target with multiple facilities); the total number of stars with spectroscopy is 282 of our original 286 featureless star candidates. In the following subsections we describe the observational setup used for each of these facilities.

\subsection{Shane/Kast observations}

We obtained spectroscopic observations of 82 candidate featureless stars using the Kast spectrograph on the Shane 3\,m telescope at Lick Observatory during semesters 2023A to 2024B, for a total of 15 nights of observations. Our goal was to obtain a signal-to-noise ratio S/N>20 per 1\,\AA\ pixel. Kast is a dual channel spectrograph, consisting of a user-selected dichroic that separates a blue and red arm. During the first three nights of observations, we used the d46 dichroic, combined with the 830/3460 grism (0.94\AA/pixel) for the blue arm and the 1200/5000 grating (0.65\AA/pixel) for the red arm. The wavelength coverage of this setup is $3400-4500$\,\AA\ and $4750-6300$\,\AA. For all subsequent observations, we instead used the d55 dichroic combined with the 600/4310 grism (1.02\AA/pixel) on the blue side and the 600/7500 grating (1.3\AA/pixel) on the red side. This setup allows us to cover a much broader wavelength range ($3500-5400$\,\AA\ and $5800-9000$\,\AA) without compromising too heavily on the instrument resolution ($R\sim1700$ and $R\sim2400$ for the blue and red arms, respectively). Individual exposures were typically 1200\,s long, with at least 3 exposures per target. Additional exposures were acquired to reach our requisite S/N.

\subsection{SOAR/Goodman observations}

We observed 69 candidate featureless stars with the SOAR/Goodman spectrograph during semesters 2023A and 2023B. The data were collected in service mode, using the \texttt{SYZY\_400} grating with a $1$ arcsec slit, and $2\times2$ detector binning. This setup delivers a spectral resolution $R\sim850$ at 5500\,\AA, and wavelength coverage $3300-7100$\,\AA, covering the most important optical transitions of white dwarf stars. We observed each star for a typical integration of $2\times300\,{\rm s}$, while the faintest stars in our sample were observed with either $2\times600\,{\rm s}$ or $4\times600\,{\rm s}$ to meet the requisite S/N>20 per 1\,\AA\ pixel.

\subsection{Keck/DEIMOS observations}

We observed 13 candidate featureless stars using Keck/DEIMOS as a backup programme during an observing run on 2022 July 26-27. Most of the data were collected during morning twilight with a 2 arcsec slit in good (sub-arcsecond) conditions. We used the 600ZD grating combined with the GG455 filter with $1\times1$ on chip binning. We obtained a single 300\,s exposure of each target. The total wavelength coverage of the observations is $4900-10000$\,\AA, with a small gap near $7500$\,\AA\ due to the gap between adjacent detectors. The FWHM resolution of the data are $\sim 4$\,\AA\ for a uniformly illuminated slit. Given the seeing conditions during the observations ($\lesssim 1''$), the resolution is likely somewhat better than this ($\lesssim2\,$\AA).

\subsection{VLT/X-shooter observations}
\label{sec:xshooter}

We conducted our VLT/X-shooter observations as a snapshot ``any weather'' programme during Period 111 (Programme ID: 111.24LB.001). Each candidate was observed with flexible observing conditions, with constraints on the airmass ($< 2$) and seeing ($<1.5''$), which was often well-matched to the adopted slit widths (UVB=$1.3''$, VIS=NIR=$1.2''$). Each candidate was observed for just one exposure, with an exposure time corresponding roughly to a ${\rm S/N} \gtrsim 15$ per $12~{\rm km~s}^{-1}$ pixel. The UVB and VIS data were both binned $2\times2$ on-chip, and deliver a full width at half maximum (FWHM) spectral resolution of $v_{\rm FWHM} \simeq 65~{\rm km~s}^{-1}$, corresponding to $R\simeq4600$. We observed a total of 174 featureless star candidates. Note that the X-shooter data reported in this paper are of the highest spectral resolution and S/N of all our samples, and are the main focus of this work. We also note that any featureless candidates that were identified by any of our other observing programmes were also observed with X-shooter where possible. Finally, we also obtained deep X-shooter observations of four blackbody stars that were reported by \citet{SuzukiFukugita2018}: J0027$-$0017, J0146$-$0051, J1255$+$1924, and J1343$+$2706.

\subsection{Data reduction}

All data were uniformly reduced with the \texttt{PypeIt} data reduction pipeline \citep{PypeIt2020}. This software subtracts the overscan regions, trims and orients the data, traces the slit edges, corrects for the pixel-to-pixel sensitivity variations and corrects for the spatial illumination profile along the slit. Wavelength calibration is performed relative to a reference calibration frame, depending on the instrument, and regions of constant wavelength are mapped over the entire detector by tracing along the ridges of the arc calibration emission lines. Objects are automatically identified and traced along the spectral direction, and the corresponding sky background regions are identified. A global model of the sky background emission is constructed, in addition to a local sky background emission in the immediate vicinity of each object trace. The objects are extracted using both an optimal and a boxcar algorithm. The data are flux-calibrated relative to a standard star that was often acquired during the twilight hours of the same night (or as near as possible). As our observing programme progressed, we instead used our most featureless and brightest stars to flux calibrate the data. This is particularly helpful to mitigate spurious features being introduced in the sensitivity function near the absorption line features that are present in the more commonly used standard stars. As a final step, stars with more than one exposure acquired with the same telescope and setup were coadded into a single spectrum using the standard \texttt{PypeIt} routines. No corrections were made to account for slit losses.

\section{Analysis}
\label{sec:analysis}

The goal of this paper is to identify new featureless white dwarf spectra that are consistent with a blackbody functional form. Such stars are among the DC white dwarf stars (see \citealt{Blouin2024} for a recent overview of white dwarf classification). White dwarfs are usually classified according to the system proposed by \citet{Sion1983}, where DC white dwarfs are objects with absorption or emission lines that are within 5 percent of the continuum level. Since this metric depends on the spectral resolution of the data, we have instead adopted an approach that depends on the measured equivalent width (EW; or a corresponding $2\sigma$ upper limit) in the vicinity of certain spectral features. This has allowed us to quantify the strength of individual absorption features, independent of the spectral resolution and S/N of the data. From this classification,  we identify the subset of spectra that appear to be featureless based on a limiting EW.

\begin{figure}
	\includegraphics[width=\columnwidth]{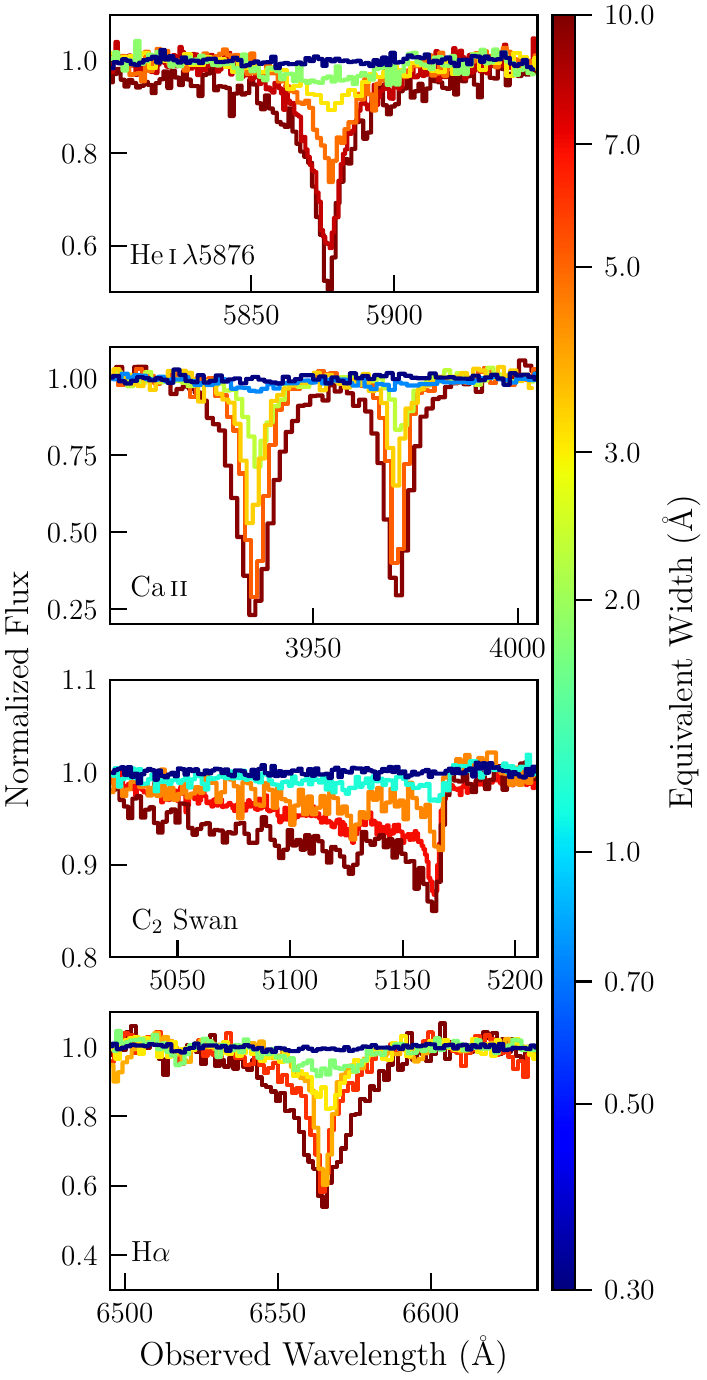}
    \caption{Example spectra of our candidate featureless white dwarf stars, color-coded by the equivalent width of the absorption line feature (see colorbar). From top to bottom, we show the absorption features \HeI\,$\lambda5876$, the \CaII\,$\lambda\lambda3933,3968$ doublet, the ${\rm C}_{2}$ Swan band, and \Ha.}
    \label{fig:examplespec}
\end{figure}

The key absorption lines that we focus on in this paper include \Ha, \HeI\ 5876\,\AA, the \CaII\ doublet $\lambda\lambda$3933,3968,\,\AA, and the ${\rm C}_{2}$ Swan band at $\sim5167$\,\AA. For each of the stars observed by our programme, we fit a low order polynomial to the continuum using regions deemed to be free of absorption near the aforementioned features. We then calculated the EW and its corresponding uncertainty. To provide a visual guide to the line strength for a given equivalent width, we show a collection of real spectra in Figure~\ref{fig:examplespec} that are colour-coded by the EW of the absorption line.

\begin{figure}
	\includegraphics[width=\columnwidth]{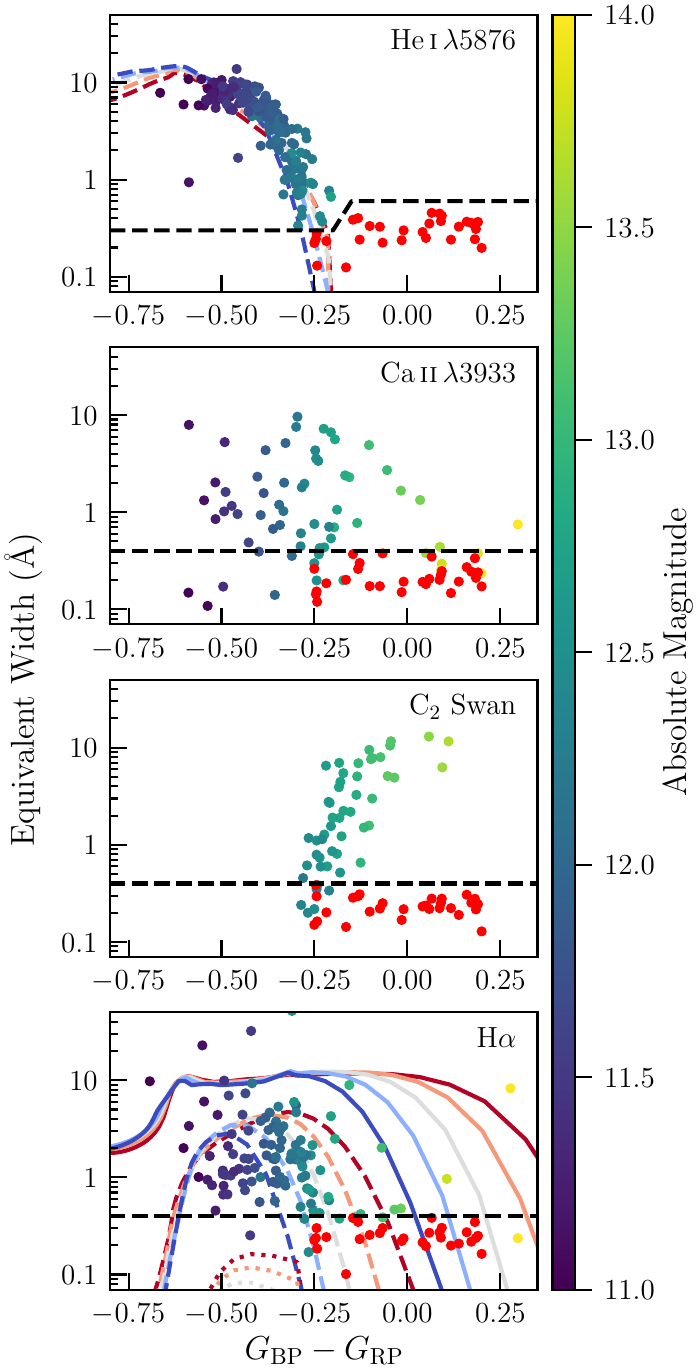}
    \caption{Equivalent widths of the four diagnostic absorption lines (see panel labels) used in our study as a function of the \emph{Gaia} BP/RP colour. Absorption lines that are detected at $>3\sigma$ confidence are shown by the blue/green/yellow symbols, colour-coded by the absolute magnitude. The red symbols show $3\sigma$ upper limits on features that are undetected. We adopt a $3\sigma$ limiting equivalent width of 400\,m\AA\ for \CaII\,$\lambda3933$, C$_{2}$ Swan, and \Ha, while we use a $3\sigma$ limiting equivalent width of 300\,m\AA\ for \HeI\,$\lambda5876$ (as indicated by the horizontal black dashed line in all panels). Given the strong trend of \HeI\ in the top panel, we also assume that all stars with $G_{\rm BP}-G_{\rm RP}\geq-0.20$ exhibit \HeI\,$\lambda5876$ absorption weaker than our 300\,m\AA\ cutoff. The coloured curves shown in the top and bottom panel are based on the 1D DBA/DB/DC models of \citet{Cukanovaite2021}, where the colour of the curve represents a surface gravity of $\log\,g=7.0,~7.5,~8.0,~8.5,~9.0$ (expressed in cgs units, from red to blue). The linestyle represents a He/H abundance of $10^{2}$, $10^{5}$, and $10^{8}$ (solid, dashed, dotted, respectively). Note that each curve represents a range of atmosphere temperatures; the model temperature is anti-correlated with the \textit{Gaia} colour. In the top panel, we only display the ${\rm He/H}=10^{5}$ models, since all models are qualitatively similar.}
    \label{fig:EWcolmag}
\end{figure}

In Figure~\ref{fig:EWcolmag}, we present the derived EWs of each line as a function of the Gaia $G_{\rm BP}-G_{\rm RP}$ colour, where each point is colour-coded by the absolute magnitude of the white dwarf. Several interesting features are observed in these plots. First, the EW of the \HeI\ absorption features are stronger for bluer and intrinsically brighter white dwarfs. The strongest optical \HeI\ line (at $\lambda5876$\,\AA) becomes extremely weak at cooler temperatures (redder colours) when the Gaia colour exceeds $G_{\rm BP}-G_{\rm RP}\gtrsim-0.20$. Around this colour, some of the white dwarfs show detectable C$_{2}$ Swan absorption, with the strength increasing towards red colours. Most of the stars that we have observed either have detectable
\HeI\ or C$_{2}$ Swan absorption, but it is rare that both features are seen in the same white dwarf spectrum. The remaining stars occasionally show \CaII\ or H Balmer absorption, but this generally occurs at blue colours. Also shown in Figure~\ref{fig:EWcolmag} are the DBA/DB/DC 1D white dwarf models of \citet{Cukanovaite2021}.\footnote{We retrieved these model data from: \url{https://warwick.ac.uk/fac/sci/physics/research/astro/people/tremblay/modelgrids/}} These models show a good overall agreement with the equivalent widths measured for our white dwarf sample. We also note that the 3D models calculated by \citet{Cukanovaite2021} are qualitatively similar to the 1D models. In our work, we adopt the 1D models, since these extend to much cooler effective temperatures ($\sim3500~{\rm K}$) relative to the 3D models ($\sim12000~{\rm K}$); the cooler temperatures (and redder colours) of the 1D models are more consistent with our featureless star sample.

Based on the qualitative examples of the spectra shown in Figure~\ref{fig:examplespec}, and the quantitative EWs in Figure~\ref{fig:EWcolmag}, we impose a rather strict criteria to define featureless white dwarf stars:
(i) the \Ha, \CaII\,$\lambda3933$, and C$_{2}$ Swan absorption lines must have a measured ${\rm EW}<0.4$\,\AA;
(ii) the \HeI\,$\lambda5876$\,\AA\ absorption line must have a measured ${\rm EW}<0.3$\,\AA, or a Gaia colour $G_{\rm BP}-G_{\rm RP}\geq-0.20$. Based on this selection, our observations reveal 29 bright featureless white dwarf stars that we use in the following analysis. These stars are colour-coded with red symbols in Figure~\ref{fig:EWcolmag}. The top panel of Figure~\ref{fig:EWcolmag} demonstrates that featureless stars are only present in our sample at cooler temperatures (redder colours), and are likely an extension of the DB white dwarfs \citep{SerenelliRohrmannFukugita2019}.

We also remark on several white dwarfs in the top panel of Figure~\ref{fig:EWcolmag} that appear to be significantly below both the white dwarf models and the envelope of other white dwarf stars (at EW[\HeI\,$\lambda5876]<2$\AA\ and $G_{\rm BP}-R_{\rm RP}<-0.4$). We confirmed that the spectra of these stars appear to have genuinely low \HeI\ equivalent widths for their measured colours. We also note that the white dwarf with EW(\HeI\,$\lambda5876$) $\simeq1$\AA\ and $G_{\rm BP}-R_{\rm RP}\simeq-0.60$ exhibits one of the strongest \CaII\,$\lambda3933$\,\AA\ equivalent widths in our sample ($\sim8$\,\AA), and moderate \Ha\ absorption, so this is not a DB white dwarf star.

As an interesting side note, one of the featureless stars (J1218$+$4148) identified by \citet{SuzukiFukugita2018} has a Gaia (AB) colour $G_{\rm BP}-R_{\rm RP}=-0.67$ and no apparent \HeI\,$\lambda5876$\AA\ absorption. We therefore suppose that there might be a rare population of featureless white dwarf stars at hotter temperatures (bluer colours) that exhibit weaker than expected \HeI\ features. Furthermore, we note that this star exhibits no turnover at far UV wavelengths.

\begin{figure*}
	\includegraphics[width=0.9\textwidth]{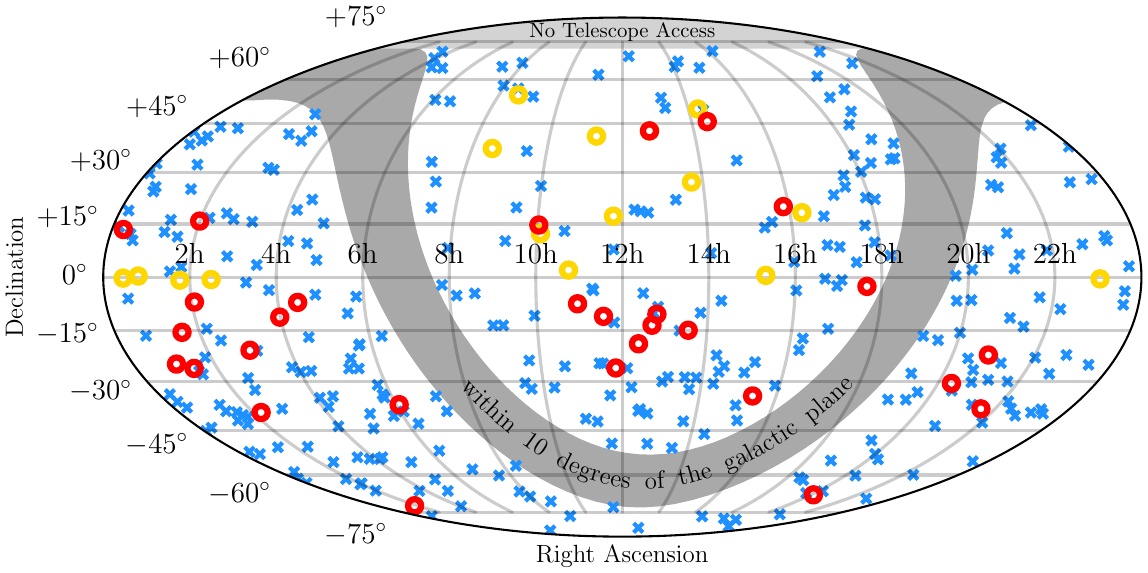}
    \caption{Sky distribution of our candidates (blue $\times$ symbols) and stars that are featureless down to a $3\sigma$ limiting equivalent width of 400\,m\AA\ (red circles; see text for further details). We also plot the \citep{SuzukiFukugita2018} blackbody star sample as yellow circles. Note that one star (J1245$+$4238) is in common between our sample and \citet{SuzukiFukugita2018}. The dark gray band marks the region within $\pm10^{\circ}$ of the Galactic plane, while the light gray region at $\delta>72^{\circ}$ marks the location that we could not observe candidates due to our telescope access.}
    \label{fig:skydist}
\end{figure*}

The sky distribution of these stars is shown in Figure~\ref{fig:skydist} (red circles) together with the other survey candidates (blue crosses). Overall, our featureless standard stars are almost uniformly distributed across the sky; most observatories worldwide will have access to at least one bright, featureless white dwarf star at any time of the year. We also show 17 \citet{SuzukiFukugita2018} blackbody stars as yellow circles. We note that our VLT/X-shooter observations of one of the \citet{SuzukiFukugita2018} stars (J1255$+$1925) revealed Ca\,\textsc{ii} absorption that does not meet our selection criteria defined above. The remaining three stars that we observed from the \citet{SuzukiFukugita2018} sample (J0027$-$0017, J0146$-$0051, J1343$+$2706) were all featureless to within the selection criteria. We also note that if any of the featureless stars in our sample exhibit an unnoticed EW=0.5\,\AA\ absorption feature at \Ha, it would change the SDSS r-band magnitude by 0.4~mmag, which is a precision of better than 0.1\%.

As a final step to check to reliability of our featureless star sample, we cross-matched each featureless star with the \emph{Gaia} catalogue to identify nearby stars that could potentially blend with and contaminate the flux of the featureless stars. We found that 22 of the 29 featureless stars have no detected sources within $10''$. Six of the remaining stars are at least $6''$ from their nearest star. Finally, there is one star in our sample (BB200707$-$673442) that is currently $4.3''$ from a nearby $G=20.4$ star. Note that all of these offsets depend on the proper motions of the stars, and users who wish to ensure the reliability of their flux calibration should ensure that proper motions do not cause future issues with blending.

\section{Blackbody parameters}
\label{sec:blackbodyparams}

The best available spectrophotometric flux standard stars are primarily based on white dwarf stars with significant absorption features. The sensitivity function of a spectrograph based on such stars often shows residuals near the strongest absorption lines (particularly the \HI\ Balmer lines). This introduces a systematic error with the calibrated flux of our candidate stars. Many of our stars were also observed in poor weather conditions, where the wavelength-dependent slit losses have not been accounted for. We are therefore unable to use our observed spectra to reliably determine the functional form of each featureless star. Instead, we have cross-matched our new sample of 29 featureless stars with wide-area survey photometry, from far-ultraviolet to mid-infrared wavelengths, accounting for the proper motions as determined from the \textit{Gaia} satellite. In what follows, we use the broad-band photometry of multiple surveys to model the flux distribution of our featureless star sample, and test the internal consistency of the various surveys. All filter profiles were retrieved from the Spanish Virtual Observatory (SVO) Filter Profile Service\footnote{The filter profiles used in this paper can be retrieved from: \url{http://svo2.cab.inta-csic.es/theory/fps/}} \citep{Rodrigo2012,Rodrigo2020}.

\subsection{Photometry}

Our measurements are all calibrated relative to the \textit{Gaia} Data Release 3 photometry and astrometry \citep{Gaia,Gaia2023}. We use the main \textit{Gaia} $G$ band magnitude as the absolute scale that determines the overall flux normalisation of the stars; every broad-band filter considered in this paper is calibrated relative to the \textit{Gaia} $G$ band. We also include measurements of the blue and red \textit{Gaia} filters, $G_{\rm BP}$ and $G_{\rm RP}$. We employ the AB zeropoint of each filter derived by \citet{Riello2021}:
\begin{eqnarray}
{\rm ZP}_{\rm AB}(G) = 25.8010 \pm 0.0028 \nonumber \\
{\rm ZP}_{\rm AB}(G_{\rm BP}) = 25.3540 \pm 0.0023 \nonumber \\
{\rm ZP}_{\rm AB}(G_{\rm RP}) = 25.1040 \pm 0.0016 \nonumber
\end{eqnarray}

The only available ultraviolet photometry currently available for our featureless star sample is the Galaxy Evolution Explorer (GALEX; \citealt{Morrissey2007}) All-sky Imaging Survey (AIS) data release GR6+7 \citep{GALEX}. This mission conducted simultaneous far-ultraviolet ($FUV$; $1344\lesssim\lambda/\textup{\AA}\lesssim1786$) and near-ultraviolet ($NUV$; $1771\lesssim\lambda/\textup{\AA}\lesssim2831$) imaging of most of the sky. We adopt an AB zeropoint of each filter ${\rm ZP}_{\rm AB}(FUV) = 18.82$ and ${\rm ZP}_{\rm AB}(NUV)=20.08$, which is based on the imaging-mode bandpass determined during ground calibration, and has been maintained throughout the entire mission. Since some DC white dwarf stars show strong \CI\ absorption \citep{Vauclair1981,Koester1982,Koester2020,Camisassa2023,Blouin2023} and far red wing \Lya\ absorption \citep{KowalskiSaumon2006} in the ultraviolet, we do not include the GALEX photometry in our fitting procedure. Instead, we extrapolate our fitting results from the optical and near-infrared to the NUV/FUV range to test if ultraviolet \CI\ or \Lya\ absorption significantly affect our featureless star sample.

We also include various optical imaging survey data. In the Northern hemisphere, we include the Sloan Digital Sky Survey (SDSS) data release 16 (DR16) $ugriz$ point spread function (PSF) photometry \citep{SDSS} and the Panoramic Survey Telescope And Rapid Response System (Pan-STARRS) data release 1 (DR1) $grizy$ photometry \citep{PS1}.
Both SDSS and Pan-STARRS intend to use the AB photometric system, although small departures from the AB system are thought to be responsible for small shifts in the $u$ and $z$ bands, such that AB$-$SDSS=$-0.04$ ($u$ band) and AB$-$SDSS=$+0.02$ ($z$ band).\footnote{For a discussion about these offsets, see \url{https://www.sdss4.org/dr16/algorithms/fluxcal/\#SDSStoAB}} Deviations from AB at the level of $1-2$ percent in all SDSS bands is reasonably expected. The Pan-STARRS photometry is somewhat better, with estimated systematic uncertainty of 0.008, 0.007, 0.009, 0.011, and 0.012 in the $grizy$ bands, respectively \citep{Chambers2016}.

In the Southern hemisphere, we use the SkyMapper Southern Sky Survey (SMSS) data release 4 (DR4) $uvgriz$ photometry \citealt{SMSS}, and the Dark Energy Survey (DES) $grizY$ photometry from the six-year calibration star catalog \citep{Rykoff2023}.\footnote{The catalog is available from the following url: \url{https://des.ncsa.illinois.edu/releases/other/Y6-standards}} Both of these optical southern sky surveys report magnitudes in the AB system. The SMSS $u$-band photometry is known to have a red leak centered on $\sim717\,{\rm nm}$ with $\sim0.7\%$ transmission relative to the centre of the bandpass, while the $v$-band has a red leak at $\sim690\,{\rm nm}$ that is an order of magnitude lower than the $u$-band leak. The zeropoint that is used for the SMSS DR4 photometry is derived from synthetic photometry of low resolution \textit{Gaia} DR3 BP/RP spectroscopy. The systematic uncertainty associated with the SMSS DR4 photometry is estimated to be $\sim0.03$ mag in the $u$ and $v$ bands, $0.01$ mag in $gri$ bands, and $0.02$ mag in the $z$ band \citep{SMSS}. The DES photometry that we use has slight deviations from the AB system ($\lesssim0.003$ mag), and a systematic uncertainty of $\sim0.012$ mag for all bands \citep{Rykoff2023}.

For the near-infrared wavelength range, we adopt the $JHK_{\rm s}$ photometry of the Two Micron All Sky Survey (2MASS) mission \citep{2MASS}. The 2MASS magnitudes are reported in the Vega system; we have converted these cataloged magnitudes to the AB system using the relationships \citep{Blanton2005}:
\begin{eqnarray}
    J_{\rm AB} = J_{\rm Vega} + 0.91 \nonumber \\
    H_{\rm AB} = H_{\rm Vega} + 1.39 \nonumber \\
    K_{\rm s,AB} = K_{\rm s,Vega} + 1.85 \nonumber
\end{eqnarray}
Note that 2MASS observed the northern and southern hemispheres of the sky separately, with some overlap. Observations of standard stars shared between both hemispheres indicate a relative photometric accuracy of better than $\sim0.02$ mag. Our featureless star sample is also useful to test the above conversion between the Vega and AB scales.

Finally, we also cross-match our featureless stars with the photometry of the $W1$ and $W2$ bands of the Wide-field Infrared Survey Explorer (WISE; \citealt{Wright2010}). We use the unWISE catalogue \citep{UnWISE}, which uses deeper imaging and accounts for sources that are partially blended at the relatively coarse resolution of WISE ($\sim6''$). To convert the unWISE Vega magnitudes to the AB scale, we use the conversion \citep{UnWISE}:
\begin{eqnarray}
    W1_{\rm AB} = W1_{\rm Vega} + 2.699 \nonumber \\
    W2_{\rm AB} = W2_{\rm Vega} + 3.339 \nonumber
\end{eqnarray}
Below, we will consider a possible mismatch between the WISE AB zero-point, and that of the optical passbands. Note that we do not include the unWISE photometry when performing our blackbody fits. Instead, we extrapolate our blackbody fits to cross-validate our model against the unWISE magnitudes, and check if any sources display a mid-infrared excess that might indicate a dusty debris disk surrounding the white dwarf \citep{Debes2011}.

\subsection{Fitting procedure}
\label{sec:photofits}

Based on our cross-matched photometry (see Supplementary data), we perform an iterative fit to simultaneously determine the model parameters that describe the flux density of our stellar sample, and assess the accuracy of the photometry. As part of this fitting procedure, we derive an estimate of the systematic bias and systematic uncertainty of each filter from many of the wide-area sky surveys.


We assume that the featureless stars are well-approximated by a Planck blackbody function, of the form:
\begin{equation}
    \label{eqn:blackbody}
    f_{\lambda} = a\frac{2hc^{2}}{\lambda^{5}}\frac{1}{\exp(hc/\lambda k_{\rm B}T)-1}
\end{equation}
where $\lambda$ is the wavelength, $h$ is the Planck constant, $c$ is the speed of light, and $k_{\rm B}$ is the Boltzmann constant. The two free parameters of this model function include a constant that determines the observed flux of the star ($a$), and the effective temperature of the white dwarf stellar atmosphere ($T$). Using all available photometry (and their reported uncertainties) for a given star, we determine the values of $a$ and $T$ that minimize the chi-squared statistic.\footnote{Two of the authors (RC and NS) independently wrote a fitting code to perform the fits. We confirmed that both codes provide consistent results.}

We also note that all of the stars in our sample are within $12-85~{\rm pc}$ of the Sun based on \emph{Gaia} parallax measurements, and are therefore within the Local Bubble. We henceforth assume that interstellar dust reddening is negligible. Furthermore, we check for circumstellar dust around our sample stars using the WISE photometry.\footnote{Soon, we will also be able to test the amount of circumstellar dust with SPHEREx \citep{Crill2020}.} We also note that the blackbody temperature has a similar dependence on wavelength in the optical as the level of extinction. Therefore, even if there is a small amount of reddening along the line-of-sight to our sample stars, the functional form of Equation~\ref{eqn:blackbody} should still capture the observed spectral energy distribution.

Following this first round of fitting, we compare the catalog photometry and errors of the $i^{\rm th}$ star to our model blackbody magnitudes ($b_{{\rm F},i}$). Each measured magnitude ($m_{{\rm F},i}$) has an associated error ($\sigma_{{\rm F},i}$) for a given filter, F. We now define the offsets from the measured value $\Delta m_{{\rm F},i}=m_{{\rm F},i}-b_{{\rm F},i}$. Now consider an unknown systematic uncertainty ($\sigma_{\rm F,sys}$) that is associated with the ground truth offset for filter F ($\Delta m_{\rm F,T}=m_{{\rm F},i,T}-b_{{\rm F},i}$), where $m_{{\rm F},i,T}$ is the true magnitude of star $i$. Since there is a systematic error associated with the true offset, any one observation of a star will deviate from the ground truth by a Gaussian random offset. Therefore, an observation might expect to measure a magnitude offset, $\Delta m_{\rm F,E}$. The probability of obtaining this expected offset is given by:
\begin{equation}
    {\rm Pr}(\Delta m_{\rm F,E}|\Delta m_{\rm F,T})=\frac{1}{\sqrt{2\pi\sigma_{\rm F,sys}^{2}}}\exp\bigg(-\frac{(\Delta m_{\rm F,E}-\Delta m_{\rm F,T})^{2}}{2\sigma_{\rm F,sys}^2}\bigg)
\end{equation}
Now, the expected offset will differ from the measured offset due to measurement error. The probability of measuring an offset $\Delta m_{{\rm F},i}$ given the expected offset is:
\begin{equation}
    {\rm Pr}(\Delta m_{{\rm F},i}|\Delta m_{\rm F,E})=\frac{1}{\sqrt{2\pi\sigma_{{\rm F},i}^{2}}}\exp\bigg(-\frac{(\Delta m_{{\rm F},i}-\Delta m_{\rm F,E})^{2}}{2\sigma_{{\rm F},i}^2}\bigg)
\end{equation}
Thus, the probability of measuring the offset $\Delta m_{{\rm F},i}$, given a ground truth offset ($\Delta m_{\rm F,T}$) with a systematic uncertainty ($\sigma_{\rm F,sys}$) for filter F is given by integrating over all possible expected offsets:
\begin{multline}
   {\rm Pr}(\Delta m_{{\rm F},i}|\Delta m_{\rm F,T})
    =\!\!\int_{-\infty}^{+\infty} \!\!\!\!\!\!\!\!{\rm Pr}(\Delta m_{{\rm F},i}|\Delta m_{\rm F,E}){\rm Pr}(\Delta m_{\rm F,E}|\Delta m_{\rm F,T}) {\rm d}\Delta m_{\rm F,E} \nonumber \\
    =\frac{1}{\sqrt{2\pi(\sigma_{{\rm F},i}^2 + \sigma_{\rm F,sys}^2)}}\exp\bigg(-\frac{(\Delta m_{{\rm F},i}-\Delta m_{\rm F,T})^{2}}{2(\sigma_{{\rm F},i}^2 + \sigma_{\rm F,sys}^2)}\bigg)
\end{multline}
To solve for the values of $\Delta m_{\rm F,T}$ and $\sigma_{\rm F,sys}$ for a given filter, we perform a joint fit to all available data (i.e. all surveys and all bands) simultaneously, and minimize the chi-squared statistic: \begin{equation}
    \chi^{2}=-0.5\sum_{i}\Big[\frac{(\Delta m_{{\rm F},i}-\Delta m_{\rm F,T})^{2}}{\sigma_{{\rm F},i}^2 + \sigma_{\rm F,sys}^2}+\log(2\pi\sigma_{{\rm F},i}^2 + 2\pi\sigma_{\rm F,sys}^2)\Big]
\end{equation}
We then apply the ground truth systematic offsets ($\Delta m_{\rm F,T}$) to the observed magnitudes, and add the systematic error ($\sigma_{\rm F,sys}$) in quadrature with the measured uncertainties, we then repeat our parametric blackbody fits to find the best-fitting $a$ and $T$ values. We iterate this process until all offsets and systematic uncertainties are unchanged between iterations, with an absolute tolerance of $10^{-4}$ magnitudes. We provide a GitHub repository with software to calculate the filter AB offsets and systematic uncertainty, as described above, allowing readers to calculate AB offsets for future photometry surveys that observe the featureless standard stars in our sample.\footnote{\url{https://github.com/rcooke-ast/AB-Filter-Correction}}

\begin{figure}
	\includegraphics[width=\columnwidth]{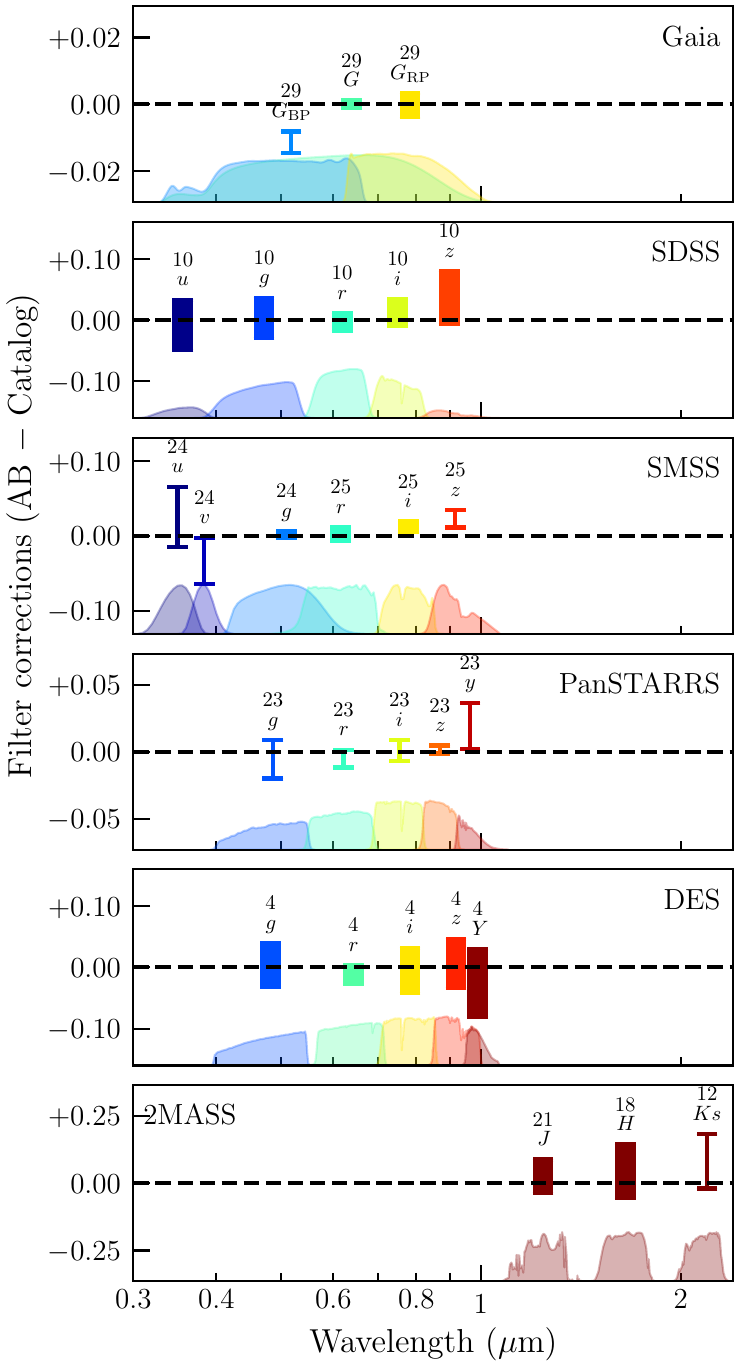}
    \caption{Each panel shows the filter corrections that need to be added to the reported catalog magnitudes to obtain AB magnitudes that are consistent with the blackbody curves we derive for our sample of featureless stars. The colour shaded regions at the bottom of each panel show each filter bandpass. 
    If the systematic uncertainty (i.e. the intrinsic dispersion $\equiv\pm\sigma_{\rm F,sys}$) is measured with at least $2\sigma$ confidence, we use thin lines with errorbars to represent the offset ($\Delta m_{\rm F,T}$) and $\pm1\sigma$ systematic uncertainty associated with each bandpass. If $\sigma_{\rm F,sys}$ is not measured with $2\sigma$ confidence we instead show vertical bands that cover the $\pm2\sigma$ range of the systematic uncertainty, centered on the measured offset. The number of featureless stars used to determine the corrections is listed above the filter label. Note the different y-axis scales used for each panel.
    The results are also compiled in Table~\ref{tab:filtercorr}.}
    \label{fig:filtercorr}
\end{figure}

Using the above formalism, we compile a list of the best-fitting corrections in Table~\ref{tab:filtercorr} that should be applied to each filter to convert the cataloged magnitudes to the AB system based on the featureless blackbody fits to our star sample. We also provide an estimate of the systematic uncertainty associated with the cataloged magnitudes. If the offset and dispersion are not inferred with $2\sigma$ confidence, we instead report a $2\sigma$ upper limit. These values are illustrated in Figure~\ref{fig:filtercorr}, where the central value of each bar represents the offset ($\Delta m_{\rm F,T}$), and the height of each thin bar represents the $\pm\sigma_{\rm F,sys}$ of each filter. When the intrinsic dispersion is not detected with $>2\sigma$ confidence, we use a thick bar and represent the height of the thick bar as the $\pm2\sigma$ upper limit on the intrinsic dispersion. For reference, the filter passbands used in our analysis are shown at the bottom of each panel.


 


\begin{table}
    \centering
    \caption{Filter Corrections (AB = Catalogue + $\Delta m_{\rm F,T}$).}
    \begin{tabular}{lcccc}
\hline
Survey + Filter & $N$ & $\Delta m_{\rm F,T}$ & $\sigma_{\rm F,sys}$ & Sig. \\
\hline
\hline
Gaia $G$ & 29 & $0.0^{\rm a}$ & $\ldots$ & $\ldots$ \\
Gaia $G_{\rm BP}$ & 29 & $-0.0114\pm0.0010$ & $0.0033^{+0.0012}_{-0.0011}$ & $10.9\sigma$ \\
Gaia $G_{\rm RP}$ & 29 & $-0.00026\pm0.00082$ & $<+0.0036$ & $0.3\sigma$ \\
\hline
SDSS $u$ & 10 & $-0.0074\pm0.0089$ & $<+0.041$ & $0.8\sigma$ \\
SDSS $g$ & 10 & $+0.0037\pm0.0082$ & $<+0.033$ & $0.5\sigma$ \\
SDSS $r$ & 10 & $-0.0030\pm0.0052$ & $<+0.014$ & $0.6\sigma$ \\
SDSS $i$ & 10 & $+0.0128\pm0.0056$ & $<+0.022$ & $2.3\sigma$ \\
SDSS $z$ & 10 & $+0.0375\pm0.0090$ & $<+0.043$ & $4.2\sigma$ \\
\hline
SMSS $u$ & 24 & $+0.0254\pm0.0097$ & $0.0402^{+0.0098}_{-0.0079}$ & $2.6\sigma$ \\
SMSS $v$ & 24 & $-0.0335\pm0.0079$ & $0.0311^{+0.0079}_{-0.0063}$ & $4.2\sigma$ \\
SMSS $g$ & 24 & $+0.0017\pm0.0015$ & $<+0.0046$ & $1.2\sigma$ \\
SMSS $r$ & 25 & $+0.0026\pm0.0020$ & $<+0.0089$ & $1.3\sigma$ \\
SMSS $i$ & 25 & $+0.0120\pm0.0017$ & $<+0.0072$ & $7.1\sigma$ \\
SMSS $z$ & 25 & $+0.0228\pm0.0033$ & $0.0115^{+0.0040}_{-0.0036}$ & $6.8\sigma$ \\
\hline
PanSTARRS $g$ & 23 & $-0.0057\pm0.0032$ & $0.0144^{+0.0029}_{-0.0023}$ & $1.8\sigma$ \\
PanSTARRS $r$ & 23 & $-0.0052\pm0.0017$ & $0.0066^{+0.0015}_{-0.0012}$ & $3.1\sigma$ \\
PanSTARRS $i$ & 23 & $+0.0008\pm0.0019$ & $0.0079^{+0.0018}_{-0.0014}$ & $0.4\sigma$ \\
PanSTARRS $z$ & 23 & $+0.0016\pm0.0010$ & $0.00297^{+0.0012}_{-0.0010}$ & $1.5\sigma$ \\
PanSTARRS $y$ & 23 & $+0.0192\pm0.0041$ & $0.0174^{+0.0038}_{-0.0030}$ & $4.7\sigma$ \\
\hline
DES $g$ & 4 & $+0.0035\pm0.0054$ & $<+0.036$ & $0.6\sigma$ \\
DES $r$ & 4 & $-0.0120\pm0.0023$ & $<+0.016$ & $5.1\sigma$ \\
DES $i$ & 4 & $-0.0045\pm0.0054$ & $<+0.036$ & $0.8\sigma$ \\
DES $z$ & 4 & $+0.0070\pm0.0060$ & $<+0.040$ & $1.2\sigma$ \\
DES $Y$ & 4 & $-0.0250\pm0.0083$ & $<+0.055$ & $3.0\sigma$ \\
\hline
2MASS $J$ & 21 & $+0.026\pm0.014$ & $<+0.062$ & $1.8\sigma$ \\
2MASS $H$ & 18 & $+0.044\pm0.024$ & $<+0.10$ & $1.8\sigma$ \\
2MASS $Ks$ & 12 & $+0.081\pm0.046$ & $0.101^{+0.059}_{-0.039}$ & $1.8\sigma$ \\
\hline
\end{tabular}

$^{\rm a}$ All reported corrections ($\Delta m_{\rm F,T}$) and intrinsic dispersions ($\sigma_{\rm F,sys}$) are quoted relative to the Gaia $G$ magnitude.
    \label{tab:filtercorr}
\end{table}

\subsection{Survey offsets from AB}
\label{sec:offsets}

The final column of Table~\ref{tab:filtercorr} lists the significance of the deviation from the AB magnitude scale (${\rm Sig}.=\Delta m_{\rm F,T}/\sigma_{\Delta m_{\rm F,T}}$). Based on our analysis, we find that most of the reported magnitudes from the wide-area photometry surveys considered here are close to the AB magnitude scale to within $2\sigma$, especially if we take into account the estimated intrinsic dispersions.

The most significant offset we find is for the Gaia $G_{\rm BP}$ filter ($10.9\sigma$) with an offset of $\Delta m_{\rm F,T}=-0.0114\pm0.0010$. Our estimated intrinsic dispersion 0.0033 mag agrees well with the official \emph{Gaia} survey zeropoint uncertainty \citep[0.0023 mag;][]{Riello2021}. Meanwhile, we find that the Gaia $G_{\rm RP}$ is within 0.001 of AB, which is somewhat better than the \textit{Gaia} zeropoint uncertainty \citep[0.0016 mag;][]{Riello2021}. We therefore conclude that the catalogued Gaia $G_{\rm BP}$ magnitudes should be reduced by 0.0114 mag to bring all Gaia magnitudes onto the AB magnitude scale.

We estimate a typical 1-2 percent uncertainty for all SDSS bands, in good agreement with the reported photometric uncertainties \citep{SDSS}. However, the offsets relative to AB that we derive are different from those quoted by the SDSS collaboration. The largest difference is the $u$ band, which the SDSS collaboration report catalogued magnitudes that are $0.04$ mag too high. Contrary to this, we find $u$, $g$ and $r$ bands to be close to AB, with the catalogued $i$ magnitudes being $\sim 1$ percent lower than AB, while the $z$ band shows a bias of $0.0375$ mag below AB (compared to $0.02$ mag reported by the SDSS collaboration). AB offsets for the SDSS survey were also estimated by \citet{Scolnic2015},\footnote{Note that \citet{Scolnic2015} define their AB offsets opposite to our definition. The values given in the penultimate column of their Table 1 are therefore the negative of our offsets.} with values $\Delta m_{\rm g,T}=+0.028$, $\Delta m_{\rm r,T}=+0.014$, $\Delta m_{\rm i,T}=+0.027$, and $\Delta m_{\rm z,T}=+0.020$. These corrections are about 17 mmag larger than those that we infer, with the exception of the z-band offset, which is about 17 mmag smaller than our correction. We note that all of these differences are consistent within our inferred upper limits on $\sigma_{\rm F,sys}$.

The SMSS $g$ and $r$ photometry is close to the AB scale, in good agreement with the results reported by \citet{SMSS}, with an intrinsic scatter $\lesssim0.01$ mag. We report the $i$ and $z$ band to be significantly ($7\sigma$) lower than AB scale by $\sim0.01$ mag and $\sim0.02$ mag respectively, which agrees well with the expected uncertainty \citep{SMSS}. The $u$ and $v$ bands differ from the AB scale by up to $\sim0.03$ mag, which agrees with the uncertainties reported by the SMSS collaboration  (0.03 mag) for these bands.

Based on our analysis, we find that all of the PanSTARRS bands are very close to AB, with a maximum deviation of 0.02 mag in the $Y$ band. We find good agreement with the systematic uncertainties of the photometry in the $r$ and $i$ bands reported by \citet{Chambers2016}, while our $g$ and $y$ band uncertainties are $\sim0.005$ mags larger. Our $z$ band intrinsic dispersion ($\sigma_{\rm F,sys}=0.00297$) is a factor of $\sim4$ lower than the PanSTARRS collaboration value (0.011). Furthermore, the intrinsic systematic uncertainties that we report are somewhat lower than those reported by \citet{Magnier2020}. \citet{Scolnic2015} report AB magnitude offset values in the range +0.023 to +0.033, while our estimates indicate absolute corrections of $<0.006$ for these bands. Overall, these offsets differ by $20-40$ mmag, which is significantly outside the range allowed by our inferred intrinsic systematic uncertainty of each filter $\sigma_{\rm F,sys}$. We also note that our $y$ filter offset of +0.0192 is similar to the offset reported by \citet{Magnier2020}, +0.011, which is consistent within our inferred $\sigma_{\rm y,sys}$ for this band.

\begin{figure*}
	\includegraphics[width=\textwidth]{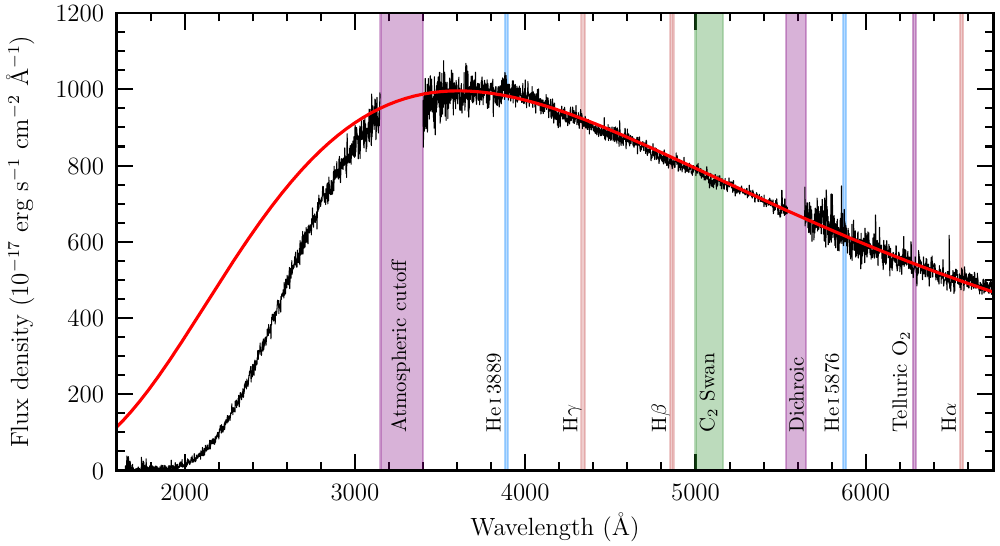}
    \caption{\emph{HST}/STIS and VLT/X-shooter spectrum of BB105735$-$073123 (black) overlaid with the blackbody fit (red) determined exclusively from the photometry of this star. The STIS data are below the atmospheric cutoff ($\lesssim3100$\,\AA), and we only show the UVB and VIS arms of X-shooter (separated by a dichroic, as indicated). To account for the slit losses, we have applied a constant scale to the X-shooter data to approximately match the blackbody curve, while no scaling correction is applied to the STIS data. The vertical bands mark the wavelengths of commonly observed telluric and white dwarf spectral features. Note that all of the \citet{Cukanovaite2021} white dwarf models are qualitatively similar to the red curve when $\log_{10}({\rm He/H})\ge5$, while the $\log_{10}({\rm He/H})=2$ model shows significant Balmer absorption, and some additional opacity in the UV that could be explained by far red wing \Lya\ absorption \citep{KowalskiSaumon2006}.}
    \label{fig:BB1057}
\end{figure*}

\begin{figure}
	\includegraphics[width=\columnwidth]{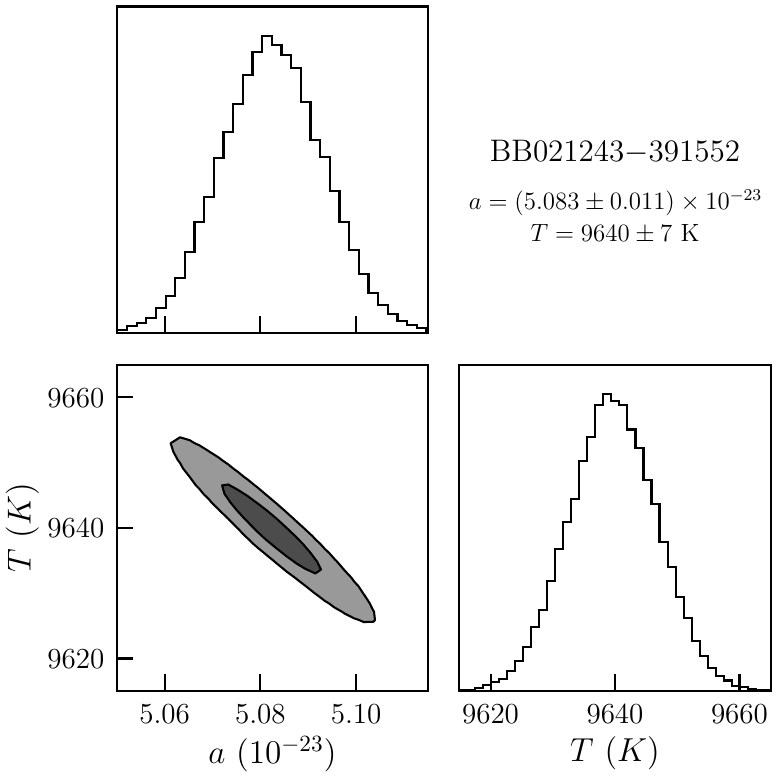}
    \caption{A corner plot illustrating the posterior distributions of the flux constant ($a$) and blackbody temperature ($T$) that is fit to the photometry measurements of BB021243$-$391552. Note that there is a significant covariance between $a$ and $T$. The reported uncertainties on these parameters do not account for the covariance between these parameters.}
    \label{fig:corner}
\end{figure}

The DES survey currently has limited coverage of our featureless star sample, with photometry of just four stars in all five bands. The currently available data suggest that DES is within 0.01 mag of AB in all bands except $Y$ which is reported to be 0.025 mag higher than AB. We currently find no evidence (within $2\sigma$) of an intrinsic dispersion among the DES bands, which is consistent with the expected intrinsic uncertainty of 0.012 mag for all bands \citep{Rykoff2023}. If the future DES footprint covers a wider area, and includes additional stars within our sample, we can reassess these AB corrections.

Although we have good statistics for 2MASS, the offsets are all consistent with AB to within $2\sigma$, and we therefore report good agreement with the conversion from Vega to AB reported by \citet{Blanton2005}. While our sample does not allow us to infer systematic uncertainties for the $J$ and $H$ bands, our upper limits agree with the 0.02 mag uncertainty that has been calculated for pairs of standard stars that have been observed in both the North and South hemispheres. We do notice a significant intrinsic dispersion for the $Ks$ band, $\sigma_{\rm F,sys}=0.101^{+0.059}_{-0.039}$, which is a factor of $\sim5$ larger than the expected 0.02 mag uncertainty.

\subsection{Flux density parameters}

Using our inferred photometric offsets and systematic uncertainties, we performed a final photometric fit to determine the best-fitting flux constant ($a$) and temperature ($T$) of each featureless star (see Equation~\ref{eqn:blackbody}). Since there is some covariance between $a$ and $T$, we use the \texttt{emcee} software \citep{emcee} to perform a Markov Chain Monte Carlo (MCMC) optimization in this final step. We setup the MCMC fit with 10 walkers initially drawn from a uniform distribution within $\pm5\sigma$ of the chi-squared minimization result. The MCMC is run for 10,000 steps with a burn in of 1000 steps and thinned by a factor of 10. We compute the integrated autocorrelation time to ensure that our chains have converged. The result for one featureless star is shown in Figure~\ref{fig:corner} using the \texttt{corner} software package \citep{corner}.

In Appendix~\ref{sec:appendix}, we show the quality of our fits to the photometry of all stars, including the residuals between the synthetic and observed photometry (data$-$model). The best-fitting values and associated uncertainties of the flux scaling constant, $a$, and the effective temperature, $T$ of each featureless star in our sample are provided in Table~\ref{tab:bbparams}, together with the original reduced $\chi^{2}$ of the fit. In the final four columns of this table, we also provide the magnitude offsets for the GALEX FUV and NUV bands, as well as the WISE W1 and W2 bands, when available. A negative value indicates that the observed magnitudes are fainter than that predicted by our blackbody fits.

We note that a significant number of our sample stars exhibit FUV and NUV fluxes that are significantly below the expected values based on our extrapolated blackbody curves. To investigate this behaviour, we cross-matched our white dwarf sample with the \emph{Hubble Space Telescope} (\emph{HST}) data archive.\footnote{\url{https://mast.stsci.edu/portal/Mashup/Clients/Mast/Portal.html}} Fortuitously, there is one white dwarf star (BB105735$-$073123) that has been observed with the Space Telescope Imaging Spectrograph (STIS; Programme ID:14076, Cycle 23, PI:B.~Gaensicke). The G230L grating was used ($R\sim500$), with an exposure time of 2293\,s, covering the wavelength range $1500-3150$\AA. We retrieved the reduced and calibrated data products from the \emph{HST} data archive, as shown in Figure~\ref{fig:BB1057}. We also plot our X-shooter spectrum of this white dwarf, flux calibrated with another featureless star, BB202025$-$302714. Due to poor observing conditions, the slit losses were considerable (see Section~\ref{sec:xshooter}). We therefore independently scaled the UVB and VIS arm data by a constant value to match the blackbody curve derived from the photometry of BB105735$-$073123 (red curve; see Table~\ref{tab:bbparams} and the supplementary information for the photometry data). We do not apply a scale correction to the \emph{HST} data. Overall, the spectral shape of the optical X-shooter data is accurately captured by the blackbody fit to the photometry. However, the blackbody curve considerably overpredicts the \emph{HST} data, becoming worse for the FUV range compared to the NUV range.

This appears to be quite common with our featureless star sample, which have FUV magnitudes considerably fainter than the expected model magnitudes, compared with the NUV magnitudes. We also note that the \emph{HST} data appear featureless to within the S/N and spectral resolution of the data. A simple blackbody functional form provides an unacceptable fit in the NUV and FUV. We explored other possible models, including the \citet{Cukanovaite2021} 1D stellar atmosphere models. All of the explored 1D white dwarf models are essentially the same as the red blackbody curve except for the models with $\log_{10}({\rm He/H})=2$, which show significant Balmer absorption, and some additional opacity in the UV. However, even the $\log_{10}({\rm He/H})=2$ models are not able to reproduce the excess UV opacity of BB105735$-$073123. Based on these \emph{HST} observations, we suppose that our featureless star sample may all continue to be featureless at UV wavelengths, but deviate from a blackbody functional form. The excess opacity could be explained by far red wing \Lya\ absorption \citep{KowalskiSaumon2006}, provided that the abundance of H in the stellar atmosphere is not capable of producing detectable (i.e. $>400$\,m\AA) Balmer absorption.

It has also been suggested that a significantly reduced FUV and NUV flux could be explained by unusually strong ultraviolet \CI\ absorption line \citep{Koester1982,Camisassa2023,Blouin2023}. Ultraviolet spectroscopy of these stars is required to determine the impact of carbon absorption in their stellar atmospheres. Currently, just one ultraviolet spectrum exists of our stars (Fig.~\ref{fig:BB1057}) and this does not show significant \CI\ absorption.

We find no significant excess flux in the WISE bands that might indicate our sample is affected by circumstellar dust, with the possible exception of BB035906$-$111835, which is strongly enhanced in the WISE $W2$ band, but not in the $W1$ band. Future observations with SPHEREx \citep{Crill2020} will allow us to further refine the possible impact of circumstellar dust on our sample. We also note that several stars show WISE fluxes that are significantly fainter compared to the blackbody fits. Although we do not have an obvious explanation for this shortfall of flux at this time, we await confirmation of these photometric measurements from SPHEREx.

Finally, we note that there are two stars that are consistent with a pure blackbody functional form from the far ultraviolet to the far infrared: BB020346$-$070136 and BB021243$-$391552. We consider these two featureless stars to follow a nearly ideal blackbody distribution.


\addtolength{\tabcolsep}{-0.3em}
\begin{landscape}
\begin{table}
    \centering
    \caption{The coordinates, proper motions, magnitude, blackbody fit parameters, and GALEX/WISE magnitude offsets of our featureless stars}
    \begin{tabular}{lcccccccccccc}
\hline
Blackbody & RA & Dec & $\mu_{\alpha}$ & $\mu_{\delta}$ & Gaia $G$ & $a$ & $T$ & $\chi^{2}/{\rm dof}$ & $\Delta{\rm FUV}^{\rm a}$ & $\Delta{\rm NUV}^{\rm a}$ & $\Delta{\rm W1}^{\rm a}$ & $\Delta{\rm W2}^{\rm a}$ \\
Star Name & (J2000) & (J2000) & $({\rm mas~yr}^{-1})$ & $({\rm mas~yr}^{-1})$ &  & ($\times10^{-23}$) & (K) & & & &  &  \\
\hline
\hline
BB001604$+$132500 & 00:16:04.221 & $+$13:25:00.645 & $+303$ & $-36$ & $17.07$ & $3.825\pm0.049$ & $9250\pm41$ & 1.107  & $\ldots$ & $\ldots$ & $-0.020\pm0.037$ & $0.02\pm0.14$ \\
BB010339$-$244456 & 01:03:39.358 & $-$24:44:56.887 & $-46$ & $-95$ & $16.81$ & $2.865\pm0.032$ & $11228\pm52$ & 0.810  & $-0.73\pm0.17$ & $-0.045\pm0.041$ & $-0.189\pm0.045$ & $\ldots$ \\
BB012524$-$260044 & 01:25:24.466 & $-$26:00:44.451 & $+235$ & $-524$ & $15.07$ & $41.43\pm0.23$ & $7798\pm14$ & 0.930  & $\ldots$ & $-0.279\pm0.021$ & $-0.1720\pm0.0074$ & $-0.070\pm0.018$ \\
BB013458$-$153608 & 01:34:58.513 & $-$15:36:08.488 & $+146$ & $+104$ & $17.09$ & $2.1486\pm0.0058$ & $11337\pm12$ & 1.536  & $\ldots$ & $\ldots$ & $-0.169\pm0.060$ & $\ldots$ \\
BB015038$-$720716 & 01:50:38.471 & $-$72:07:16.543 & $-229$ & $-232$ & $16.35$ & $15.57\pm0.19$ & $7355\pm26$ & 0.728  & $\ldots$ & $\ldots$ & $-0.054\pm0.013$ & $-0.151\pm0.036$ \\
BB015939$+$154847 & 01:59:39.366 & $+$15:48:47.628 & $-75$ & $-10$ & $15.87$ & $8.523\pm0.077$ & $10307\pm36$ & 0.522  & $-2.15\pm0.15$ & $-0.080\pm0.018$ & $-0.080\pm0.018$ & $-0.114\pm0.064$ \\
BB020346$-$070136 & 02:03:46.752 & $-$07:01:36.890 & $+33$ & $-59$ & $17.46$ & $3.233\pm0.043$ & $8709\pm39$ & 0.687  & $-0.31\pm0.37$ & $0.086\pm0.062$ & $-0.062\pm0.048$ & $\ldots$ \\
BB021243$-$391552 & 02:12:43.930 & $-$39:15:52.807 & $+11$ & $+18$ & $16.65$ & $5.083\pm0.011$ & $9640\pm7$ & 1.195  & $-0.10\pm0.12$ & $-0.035\pm0.028$ & $0.131\pm0.021$ & $0.081\pm0.078$ \\
BB030145$-$204440 & 03:01:45.684 & $-$20:44:40.133 & $+80$ & $+11$ & $17.61$ & $3.4340\pm0.0089$ & $8162\pm6$ & 1.037  & $\ldots$ & $-0.38\pm0.12$ & $0.012\pm0.043$ & $-0.18\pm0.19$ \\
BB035906$-$111835 & 03:59:06.029 & $-$11:18:35.126 & $+119$ & $-115$ & $17.54$ & $4.621\pm0.063$ & $7603\pm31$ & 1.378  & $\ldots$ & $-0.53\pm0.18$ & $0.098\pm0.036$ & $0.555\pm0.081$ \\
BB042731$-$070802 & 04:27:31.728 & $-$07:08:02.800 & $+163$ & $+40$ & $16.85$ & $9.083\pm0.080$ & $7520\pm20$ & 1.045  & $\ldots$ & $-0.275\pm0.082$ & $0.015\pm0.021$ & $0.117\pm0.064$ \\
BB060432$-$365105 & 06:04:32.032 & $-$36:51:05.882 & $-16$ & $-112$ & $17.38$ & $4.466\pm0.010$ & $8032\pm6$ & 0.552  & $\ldots$ & $\ldots$ & $-0.103\pm0.032$ & $-0.20\pm0.12$ \\
BB100149$+$144123 & 10:01:49.363 & $+$14:41:23.785 & $-338$ & $+11$ & $15.42$ & $37.25\pm0.30$ & $7316\pm18$ & 1.619  & $-1.02\pm0.48$ & $-0.065\pm0.049$ & $-0.0735\pm0.0091$ & $-0.109\pm0.024$ \\
BB105735$-$073123 & 10:57:35.134 & $-$07:31:23.176 & $-816$ & $+91$ & $14.35$ & $72.59\pm0.49$ & $8037\pm18$ & 0.463  & $\ldots$ & $\ldots$ & $-0.3086\pm0.0066$ & $-0.415\pm0.017$ \\
BB113337$-$110529 & 11:33:37.084 & $-$11:05:29.680 & $-192$ & $-59$ & $16.67$ & $3.502\pm0.039$ & $10925\pm49$ & 0.835  & $-0.55\pm0.13$ & $-0.120\pm0.028$ & $-0.170\pm0.041$ & $\ldots$ \\
BB115020$-$255335 & 11:50:20.142 & $-$25:53:35.404 & $-108$ & $-124$ & $16.40$ & $14.24\pm0.13$ & $7427\pm21$ & 0.610  & $\ldots$ & $-0.91\pm0.13$ & $-0.085\pm0.017$ & $-0.211\pm0.063$ \\
BB122313$-$185211 & 12:23:13.565 & $-$18:52:11.281 & $+24$ & $-97$ & $16.57$ & $4.971\pm0.056$ & $9942\pm42$ & 0.390  & $-1.78\pm0.29$ & $-0.042\pm0.030$ & $-0.175\pm0.034$ & $\ldots$ \\
BB124155$-$133501 & 12:41:55.920 & $-$13:35:01.266 & $-272$ & $-174$ & $15.84$ & $11.708\pm0.082$ & $9327\pm24$ & 0.875  & $\ldots$ & $\ldots$ & $-0.002\pm0.015$ & $0.141\pm0.044$ \\
BB124535$+$423824 & 12:45:35.622 & $+$42:38:24.675 & $+18$ & $-53$ & $17.20$ & $2.634\pm0.029$ & $10122\pm43$ & 0.794  & $\ldots$ & $-0.014\pm0.036$ & $-0.090\pm0.042$ & $\ldots$ \\
BB124828$-$102857 & 12:48:28.174 & $-$10:28:57.820 & $-105$ & $-95$ & $16.93$ & $6.863\pm0.056$ & $8010\pm22$ & 0.403  & $\ldots$ & $\ldots$ & $-0.802\pm0.055$ & $\ldots$ \\
BB133321$-$150026 & 13:33:21.517 & $-$15:00:26.787 & $-161$ & $-4$ & $16.95$ & $8.085\pm0.076$ & $7563\pm22$ & 0.743  & $\ldots$ & $-0.83\pm0.12$ & $-0.032\pm0.026$ & $\ldots$ \\
BB142708$+$453631 & 14:27:08.162 & $+$45:36:31.689 & $-77$ & $-34$ & $16.80$ & $7.177\pm0.063$ & $8182\pm24$ & 0.674  & $\ldots$ & $\ldots$ & $-0.042\pm0.021$ & $0.071\pm0.062$ \\
BB152316$-$341207 & 15:23:16.521 & $-$34:12:07.397 & $-364$ & $-247$ & $15.48$ & $13.76\pm0.17$ & $9868\pm42$ & 0.992  & $-1.65\pm0.40$ & $-0.102\pm0.026$ & $-0.306\pm0.017$ & $-0.273\pm0.051$ \\
BB155200$+$195753 & 15:52:00.935 & $+$19:57:53.148 & $-54$ & $-168$ & $16.30$ & $4.463\pm0.050$ & $11359\pm52$ & 0.792  & $-0.59\pm0.13$ & $-0.129\pm0.026$ & $-0.210\pm0.028$ & $-0.37\pm0.11$ \\
BB173930$-$023700 & 17:39:30.742 & $-$02:37:00.572 & $-67$ & $+204$ & $17.54$ & $3.260\pm0.033$ & $8458\pm29$ & 2.378  & $\ldots$ & $\ldots$ & $0.140\pm0.045$ & $\ldots$ \\
BB200707$-$673442 & 20:07:07.980 & $-$67:34:42.178 & $+39$ & $-151$ & $16.57$ & $8.67\pm0.10$ & $8229\pm31$ & 1.758  & $-0.67\pm0.41$ & $0.179\pm0.059$ & $-0.058\pm0.021$ & $-0.261\pm0.079$ \\
BB202025$-$302714 & 20:20:25.458 & $-$30:27:14.650 & $-71$ & $-96$ & $13.71$ & $49.64\pm0.26$ & $11240\pm26$ & 1.532  & $-0.561\pm0.038$ & $-0.165\pm0.013$ & $-0.0770\pm0.0048$ & $-0.178\pm0.013$ \\
BB205216$-$220633 & 20:52:16.841 & $-$22:06:33.421 & $+4$ & $+145$ & $15.05$ & $28.94\pm0.15$ & $8785\pm17$ & 1.118  & $\ldots$ & $-0.222\pm0.021$ & $0.0205\pm0.0077$ & $-0.087\pm0.026$ \\
BB213721$-$380838 & 21:37:21.243 & $-$38:08:38.216 & $-7$ & $-10$ & $16.65$ & $11.48\pm0.12$ & $7398\pm22$ & 0.389  & $\ldots$ & $-1.36\pm0.13$ & $0.062\pm0.018$ & $-0.039\pm0.065$ \\
\hline
\end{tabular}

    $^{\rm a}$ The final four columns of this table report the differences between the blackbody model fit to the optical/NIR data and the observed magnitude. Negative values indicate the observed magnitude is fainter than expected for the inferred blackbody distribution. The quoted error includes the observational uncertainty and the uncertainty of the model fit.

\label{tab:bbparams}
\end{table}
\end{landscape}

\begin{figure*}
	\includegraphics[width=\textwidth]{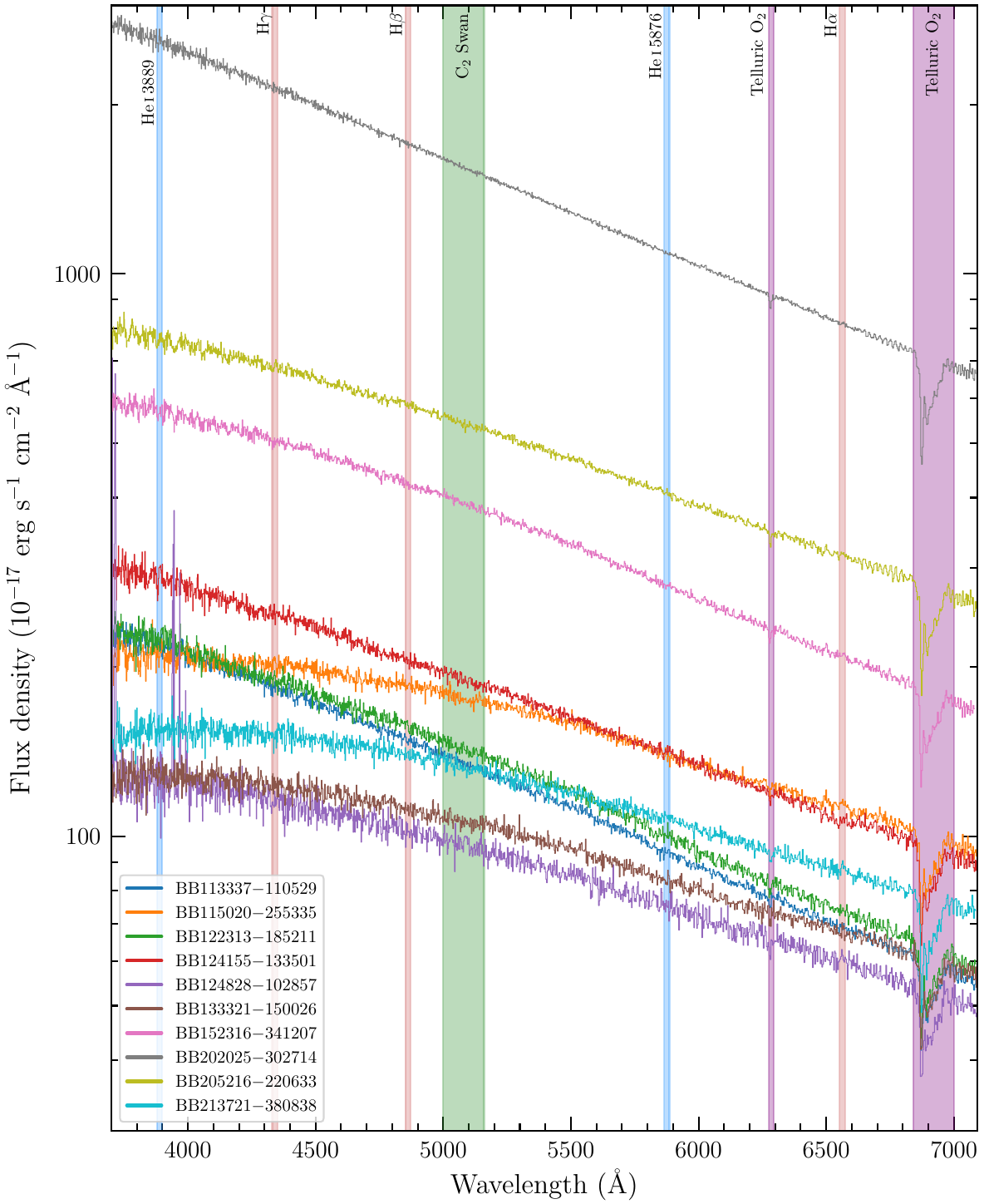}
    \caption{SOAR/Goodman spectra of 10 featureless stars in our sample. All spectra have been flux calibrated using a sensitivity function derived from the grey curve (BB202025$-$302714). The vertical bands mark the wavelengths of commonly observed telluric and white dwarf spectral features.}
    \label{fig:allBBstars}
\end{figure*}

As a final step in our data reduction, we use the best-fitting parameters of the featureless stars to recalculate the sensitivity function of our SOAR/Goodman and X-shooter data using one of our highest S/N observations of a featureless standard star (BB202025$-$302714). The resulting final calibrated and combined spectra of our sample of featureless stars are shown in Figure~\ref{fig:allBBstars}. Note that we do not use these spectra to estimate blackbody parameters; these spectra are solely illustrative to show that these stars are smooth and featureless relative to the sensitivity function derived from another featureless star.

\subsection{The future of featureless stars}

\begin{figure}
	\includegraphics[width=\columnwidth]{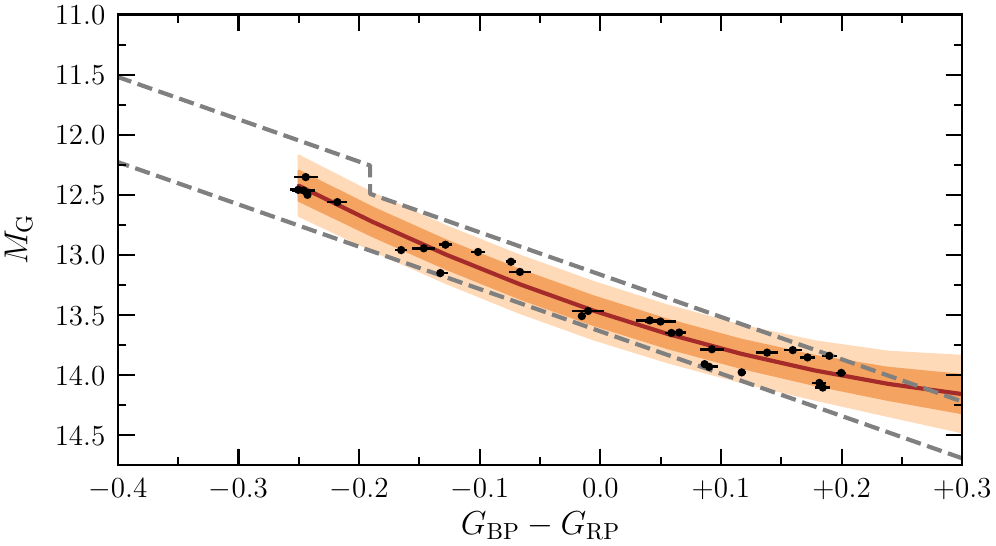}
    \caption{Our sample of 29 featureless stars (black points with error bars) are shown on the Gaia AB colour-magnitude diagram. The solid line represents a quadratic fit to the data, while the dark and light shaded regions indicate the 68 and 95 percent confidence intervals, respectively. The gray dashed line shows our original selection box (cf. Figure~\ref{fig:colmagSF18}).}
    \label{fig:colmagFinal}
\end{figure}

The primary aim of this paper is to assemble a network of the brightest and most featureless stars with an optical/NIR spectral energy distribution that is well-approximated by a blackbody functional form. This was largely based on the relatively faint featureless stars identified by \citet{SuzukiFukugita2018}. Given the brighter featureless stars that we have identified, we now reevaluate the trend seen in Figure~\ref{fig:colmagSF18} to improve the future selection of candidate featureless stars. The resulting best-fit parameters and their associated $68\%$ confidence intervals are:
\begin{eqnarray}
    \label{eqn:gaiafitb}
    \!\!\!\!M_{\rm G} \!\!\!\!&=&\!\!\!\! (13.483\pm0.033) + (3.33\pm0.15)\cdot(G_{\rm BP} - G_{\rm RP}) \nonumber \\
     &~&- (3.6\pm1.2)\cdot(G_{\rm BP} - G_{\rm RP})^{2}\\
    \!\!\!\!\sigma_{\rm int, G} \!\!\!\!&=&\!\!\!\! 0.111^{+0.018}_{-0.014}
\end{eqnarray}
where the variables take on the same meaning as in Equation~\ref{eqn:gaiafit}, and we now include a quadratic term. The resulting colour-magnitude relationship of our featureless stars is shown in Figure~\ref{fig:colmagFinal}.








\section{Conclusions}
\label{sec:conc}

We conducted a dedicated spectroscopic survey to identify bright featureless stars. 
The original goals of this paper were to:
(1) provide a definition of an optically featureless white dwarf;
(2) construct a network of featureless stars across the sky; and
(3) recalibrate the photometry of several wide-area sky surveys to the AB magnitude scale.
We draw the following conclusions:

\smallskip

(i) Based on the sample of faint SDSS blackbody stars previously identified by \citet{SuzukiFukugita2018}, we have found that featureless stars occupy a tight relationship in the Gaia colour-magnitude diagram. We exploit this tight relationship, together with low-resolution Gaia BP/RP spectroscopy, to assemble a candidate list of white dwarf stars that occupy the same space in the Gaia colour-magnitude diagram. We conducted a dedicated spectroscopic follow-up campaign of 282 stars.

\smallskip

(ii) We calculated equivalent widths of the most commonly observed features in the spectra of our white dwarf stars, including: \HeI\,$\lambda5876$, \CaII\,$\lambda3933$, ${\rm C}_{2}$ Swan, and \Ha. Our sample of stars exhibits a tight relationship between the \HeI\,$\lambda5876$ equivalent width and Gaia colour ($G_{\rm BP}-G_{\rm RP}$), as known previously, and this relationship agrees well with white dwarf models that have either a pure or predominantly helium composition. When $G_{\rm BP}-G_{\rm RP}\gtrsim-0.25$ (AB scale), the \HeI\ absorption lines become extremely weak due to the cooler stellar temperatures. Around this same colour, we identified a significant increase in the equivalent width of the ${\rm C}_{2}$ Swan absorption band.

\smallskip

(iii) In our work, we classify featureless stars as those that have a $3\sigma$ limiting equivalent width of 400\,m\AA\ for \CaII\,$\lambda3933$, C$_{2}$ Swan, and \Ha, while we use a $3\sigma$ limiting equivalent width of 300\,m\AA\ for \HeI\,$\lambda5876$. Furthermore, given the strong decline of \HeI\,$\lambda5876$ absorption at colours $G_{\rm BP}-G_{\rm RP}\gtrsim-0.25$ (AB scale), we also assume that all stars in this colour range exhibit no \HeI\,$\lambda5876$. A total of 29 stars satisfy these selection criteria. These stars occupy an approximately uniform distribution across the sky, such that any observatory worldwide should have access to at least one bright featureless star at any time of the year.

\smallskip

(iv) We model all stars with a two parameter blackbody function to determine the underlying flux density of the stars. We find that featureless stars are a subset of the cool DB stars, as previously suggested by \citet{SerenelliRohrmannFukugita2019}.

\smallskip

(v) Based on this sample, we have estimated small corrections to the AB magnitude scale of the following surveys: Gaia DR3, SDSS DR16, SMSS DR4, PanSTARRS1, DES DR6, and 2MASS. The filter corrections are compiled in Table~\ref{tab:filtercorr}. We have incorporated all of these new featureless stars into the PypeIt data reduction software, if the community wish to use them to flux calibrate their own spectroscopic data.

\smallskip

(vi) We recalculated our candidate selection algorithm based on the new discoveries reported here to enable a more efficient candidate selection in the future, when somewhat fainter featureless stars are required.

\smallskip

In the age of high-precision cosmology and flux calibration, as well as the need to find suitable photometric standard stars for the next generation of telescopes with 30+\,m aperture, we must address the current systematics associated with modeling uncertainties near the \HeI\ and Balmer absorption lines. One approach to overcome this current limitation is to identify featureless stars with a flux density that is well-approximated by a near-perfect blackbody distribution. Although such stars are rare, it is essential that we explore the possibility of using such stars to calibrate and cross-check future imaging and spectroscopic surveys, and enable sub-percent accuracy flux calibration in the optical and near-infrared wavelength range. The blackbody functional form of these stars based on optical photometry does not ubiquitously extend into the ultraviolet wavelength range, indicating that some of these stars may exhibit either far red wing \Lya\ absorption or strong \CI\ absorption lines in the ultraviolet.

\section*{Acknowledgements}
The authors would like to thank the anonymous referee for a careful and prompt review that helped to improved the clarity of several parts of this paper. During this work, RJC was funded by a Royal Society University Research Fellowship. RJC acknowledges support from STFC (ST/T000244/1, ST/X001075/1). This research has made use of the Keck Observatory Archive (KOA), which is operated by the W. M. Keck Observatory and the NASA Exoplanet Science Institute (NExScI), under contract with the National Aeronautics and Space Administration. Based on observations made with the NASA/ESA Hubble Space Telescope, and obtained from the Hubble Legacy Archive, which is a collaboration between the Space Telescope Science Institute (STScI/NASA), the Space Telescope European Coordinating Facility (ST-ECF/ESA) and the Canadian Astronomy Data Centre (CADC/NRC/CSA). This research has made use of the SVO Filter Profile Service (\url{http://svo2.cab.inta-csic.es/theory/fps/}) supported from the Spanish MINECO through grant AYA2017-84089. This work has made use of data from the European Space Agency (ESA) mission {\it Gaia} (\url{https://www.cosmos.esa.int/gaia}), processed by the {\it Gaia} Data Processing and Analysis Consortium (DPAC, \url{https://www.cosmos.esa.int/web/gaia/dpac/consortium}). Funding for the DPAC has been provided by national institutions, in particular the institutions participating in the {\it Gaia} Multilateral Agreement. This research is based on observations made with the NASA/ESA Hubble Space Telescope obtained from the Space Telescope Science Institute, which is operated by the Association of Universities for Research in Astronomy, Inc., under NASA contract NAS 5–26555. These observations are associated with program 14076. This research has made use of the SVO Filter Profile Service ``Carlos Rodrigo'', funded by MCIN/AEI/10.13039/501100011033/ through grant PID2020-112949GB-I00. This research made use of the cross-match service provided by CDS, Strasbourg.

\section*{Data Availability}

All of the data analysed in this paper are publicly available in the astronomy data archives that are linked to this paper. Our supporting research dataset, including all reduced spectroscopic data, is published in the Durham University Research Data Repository (DOI: \url{http://doi.org/10.15128/r28336h1959}). We provide the cross-matched survey and synthetic photometry of our featureless star sample as supplementary data. These supplementary data are stored as comma separated variable files, with a value of \texttt{-1.0} representing missing data for that star. Examples of how to parse these data files are provided at the following GitHub repository: \url{https://github.com/rcooke-ast/AB-Filter-Correction}



\bibliographystyle{mnras}
\bibliography{blackbody} 



\appendix

\section{Featureless star fits}
\label{sec:appendix}

\begin{figure*}
	\includegraphics[width=\textwidth]{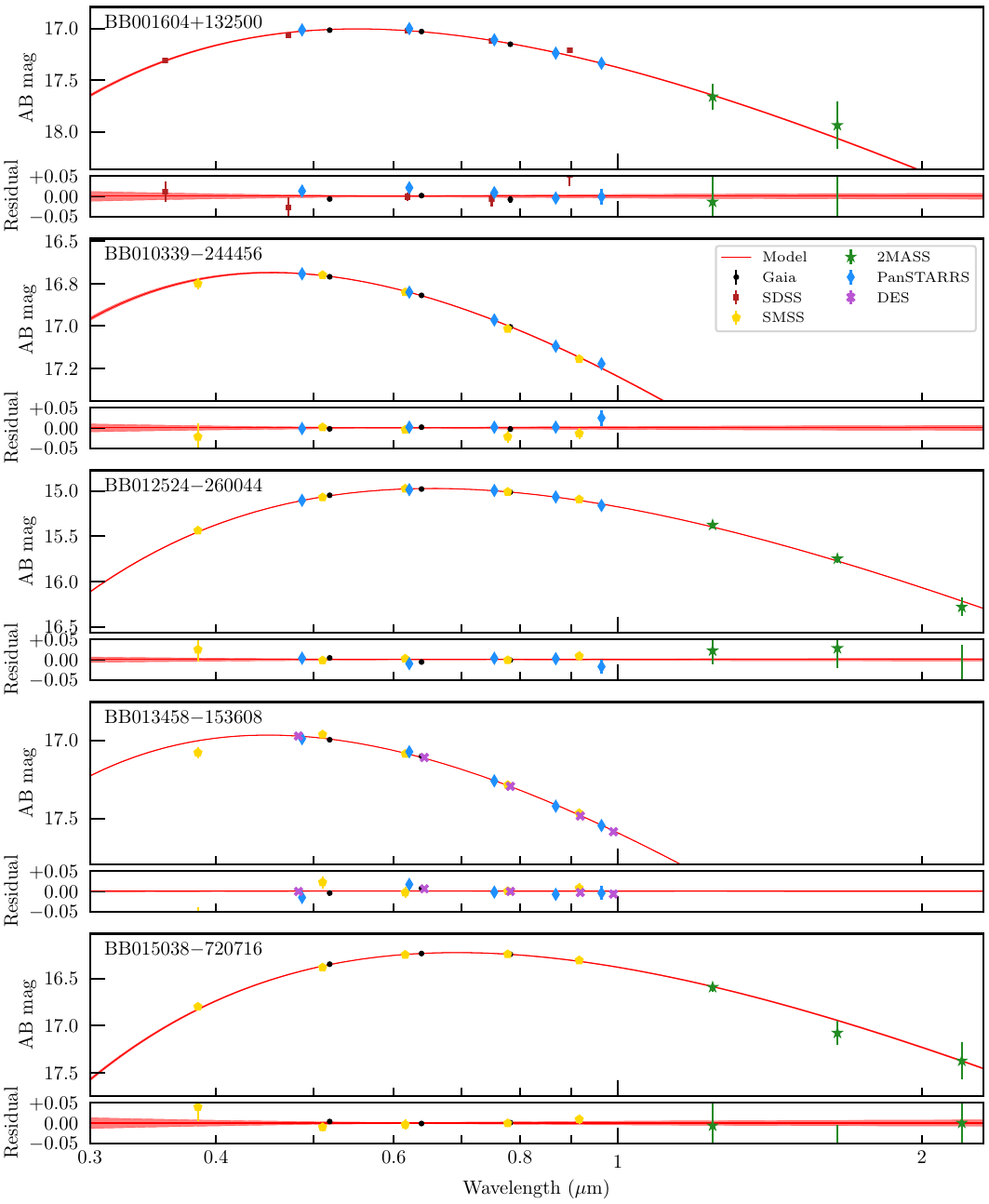}
    \caption{Blackbody fits (upper) and fit residuals (data$-$model; lower) to the photometry of our featureless stars sample. Each symbol is color-coded by the survey (see legend). The red line represents the best-fitting blackbody curve, while the red shaded region represents the $1\sigma$ confidence intervals. All of the photometry shown has been corrected to AB, based on the fitting procedure described in Section~\ref{sec:photofits} and \ref{sec:offsets}. The star name is indicated in the top left corner.}
    \label{fig:multipanelfits_1}
\end{figure*}

\begin{figure*}
	\includegraphics[width=\textwidth]{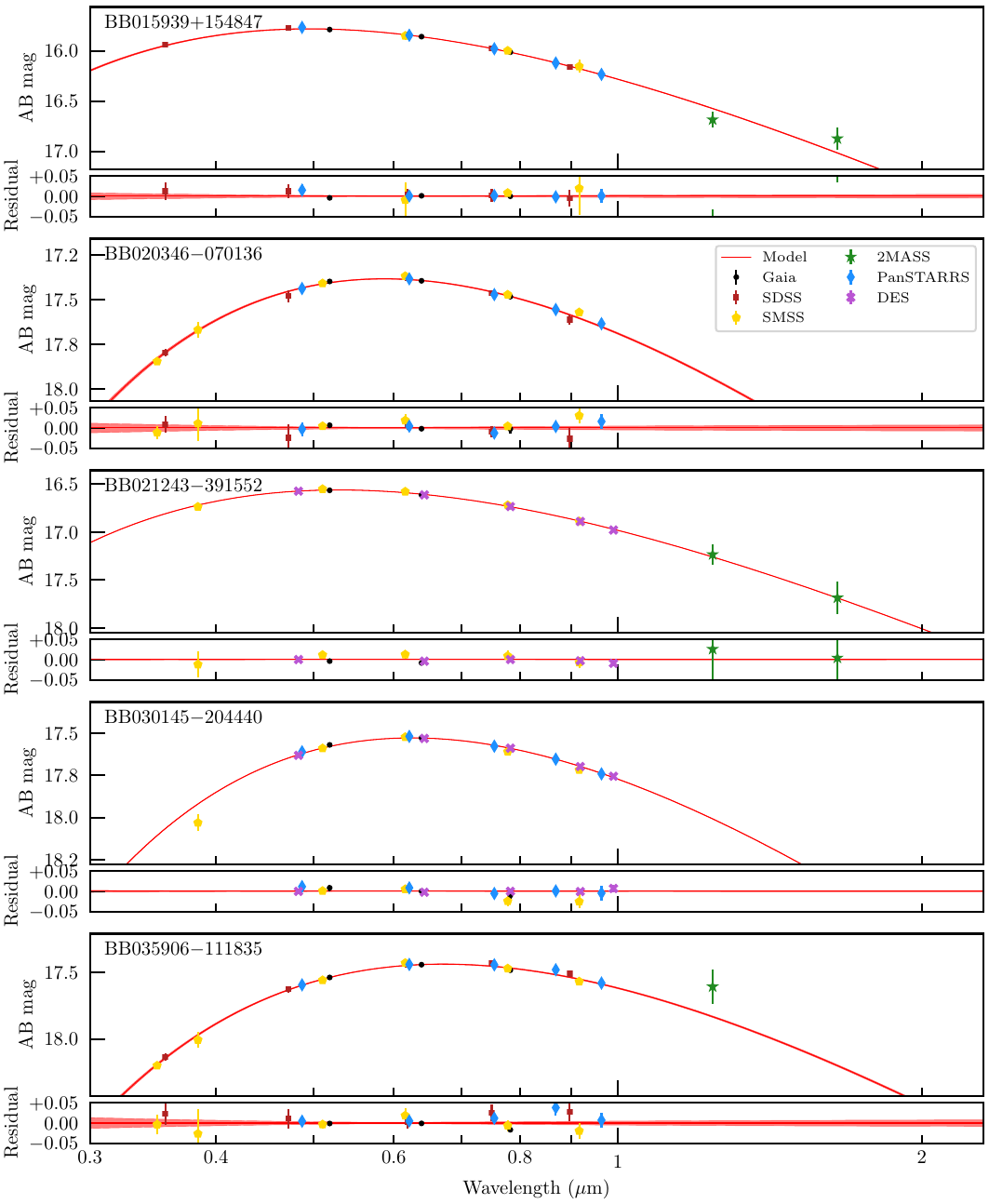}
    \caption{Same as Figure~\ref{fig:multipanelfits_1}, but showing the photometric fits to a different set of five featureless stars, as indicated in the top left corner of each panel.}
    \label{fig:multipanelfits_2}
\end{figure*}

\begin{figure*}
	\includegraphics[width=\textwidth]{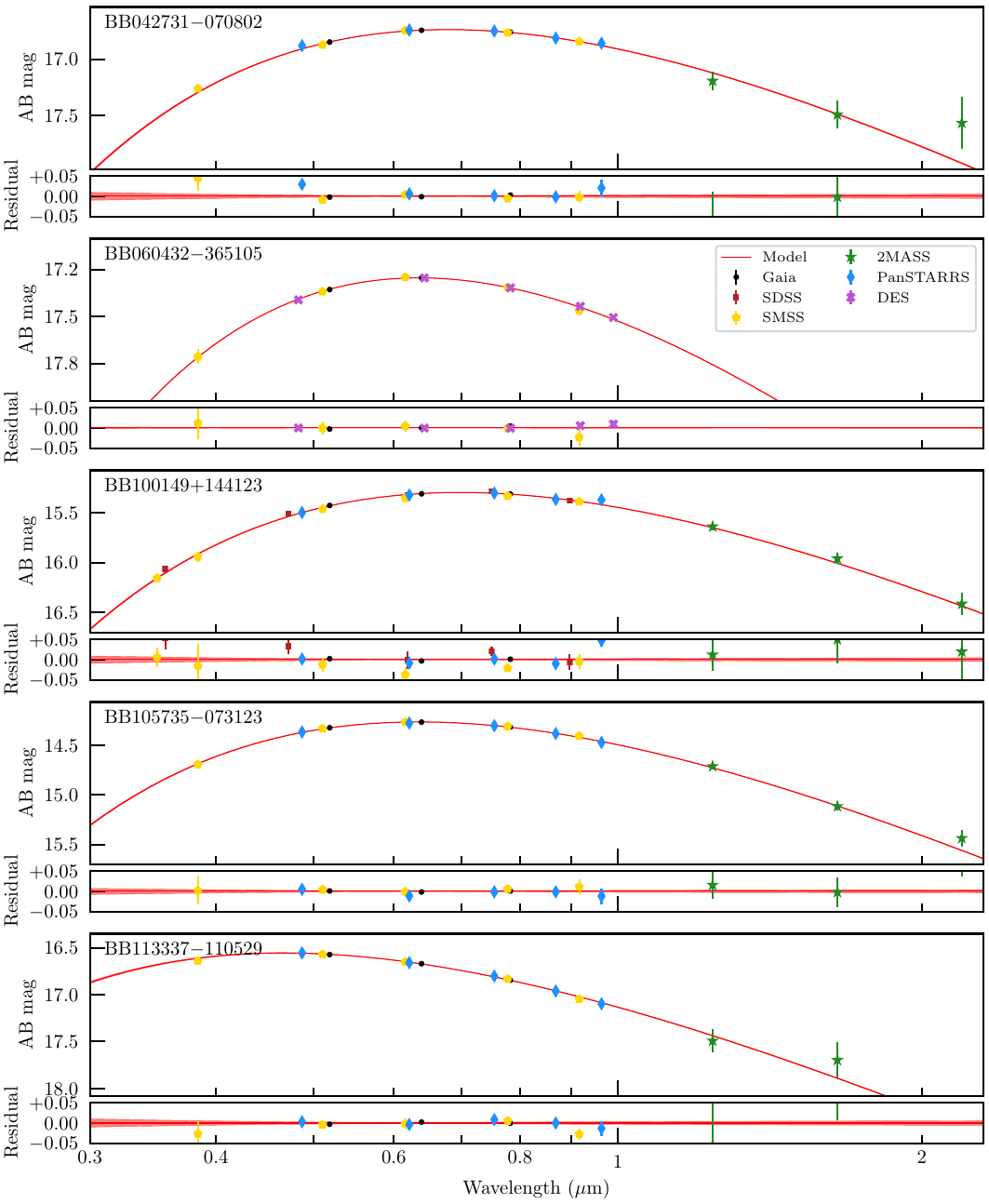}
    \caption{Same as Figure~\ref{fig:multipanelfits_1}, but showing the photometric fits to a different set of five featureless stars, as indicated in the top left corner of each panel.}
    \label{fig:multipanelfits_3}
\end{figure*}

\begin{figure*}
	\includegraphics[width=\textwidth]{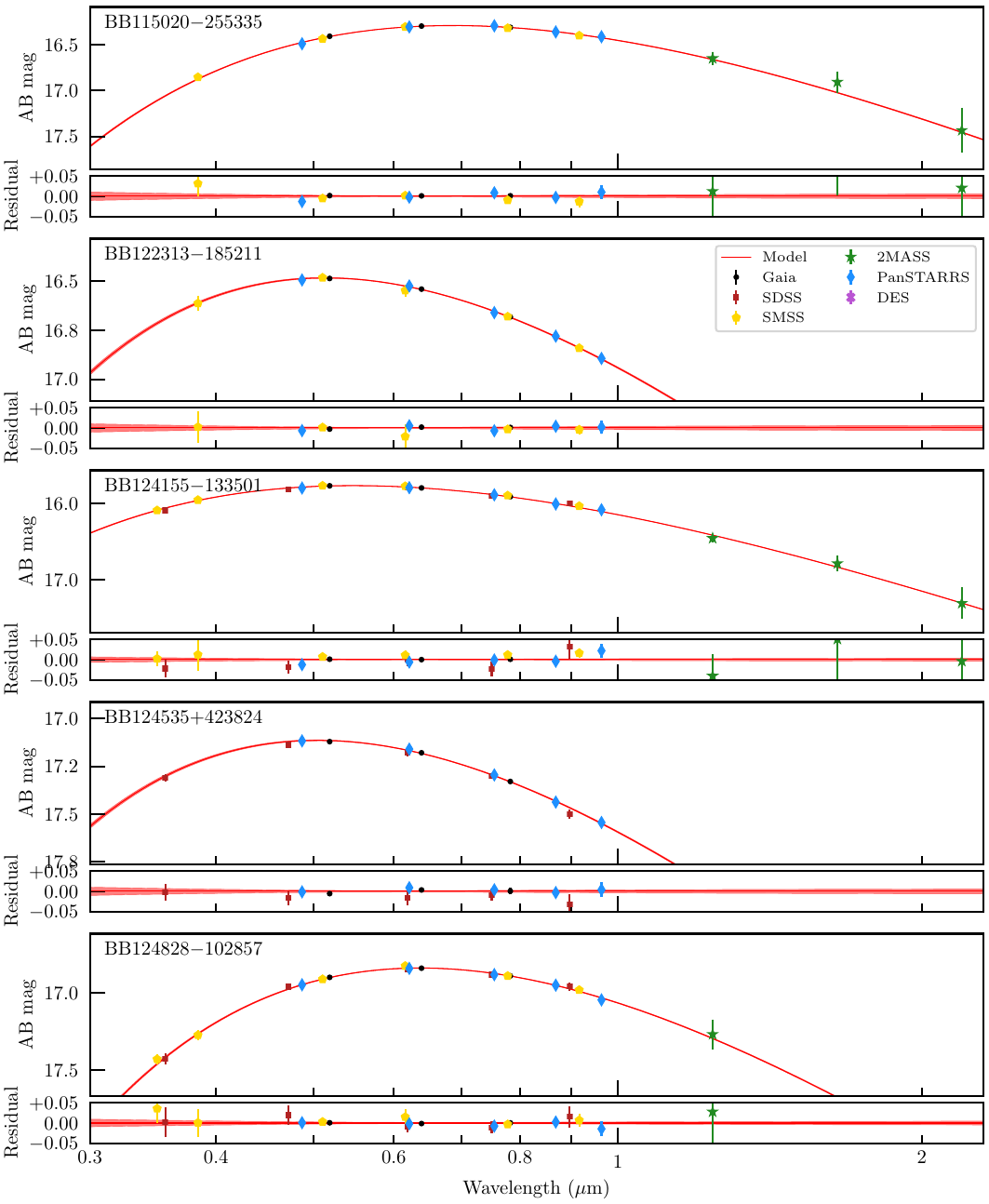}
    \caption{Same as Figure~\ref{fig:multipanelfits_1}, but showing the photometric fits to a different set of five featureless stars, as indicated in the top left corner of each panel.}
    \label{fig:multipanelfits_4}
\end{figure*}

\begin{figure*}
	\includegraphics[width=\textwidth]{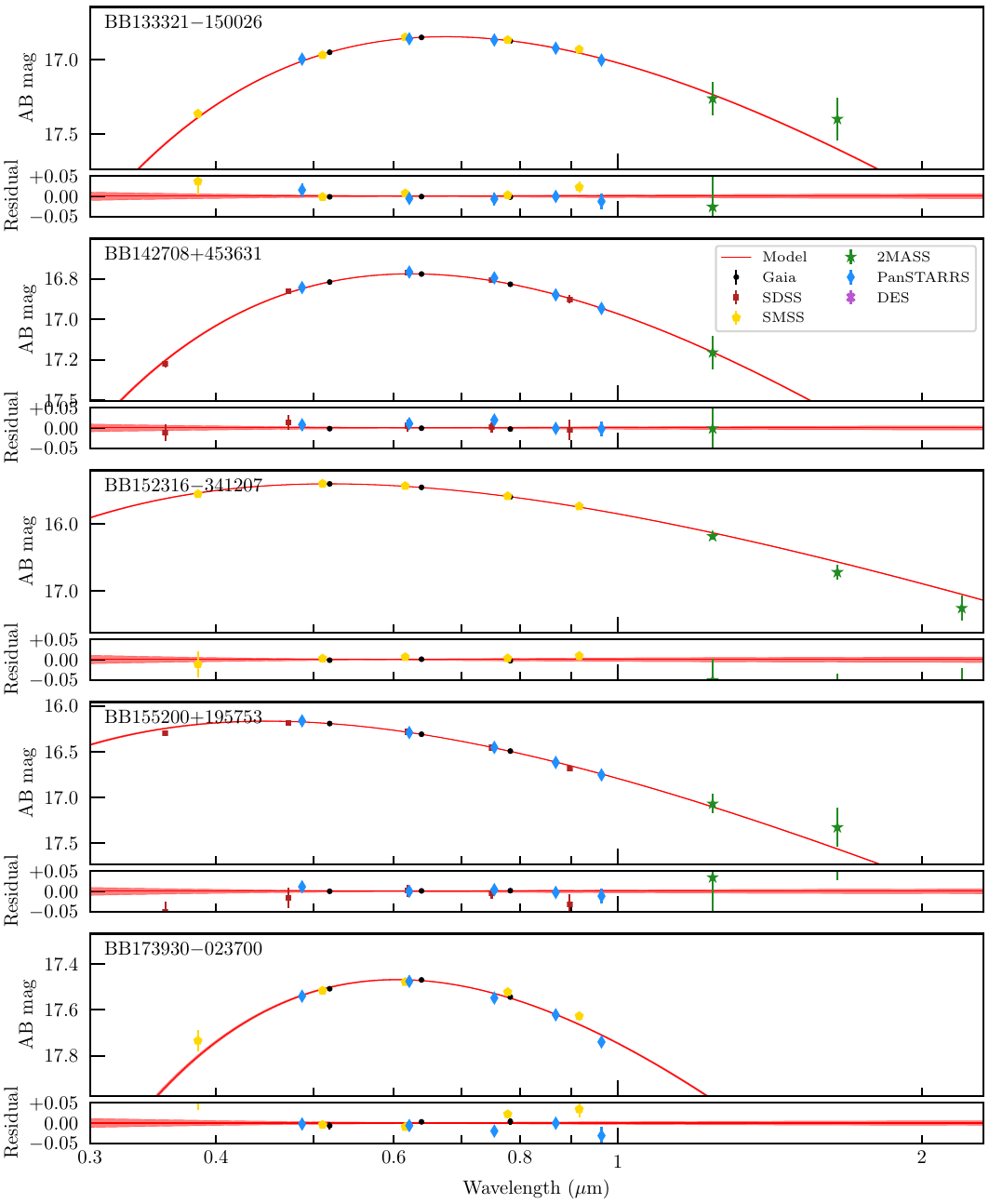}
    \caption{Same as Figure~\ref{fig:multipanelfits_1}, but showing the photometric fits to a different set of five featureless stars, as indicated in the top left corner of each panel.}
    \label{fig:multipanelfits_5}
\end{figure*}

\begin{figure*}
	\includegraphics[width=\textwidth]{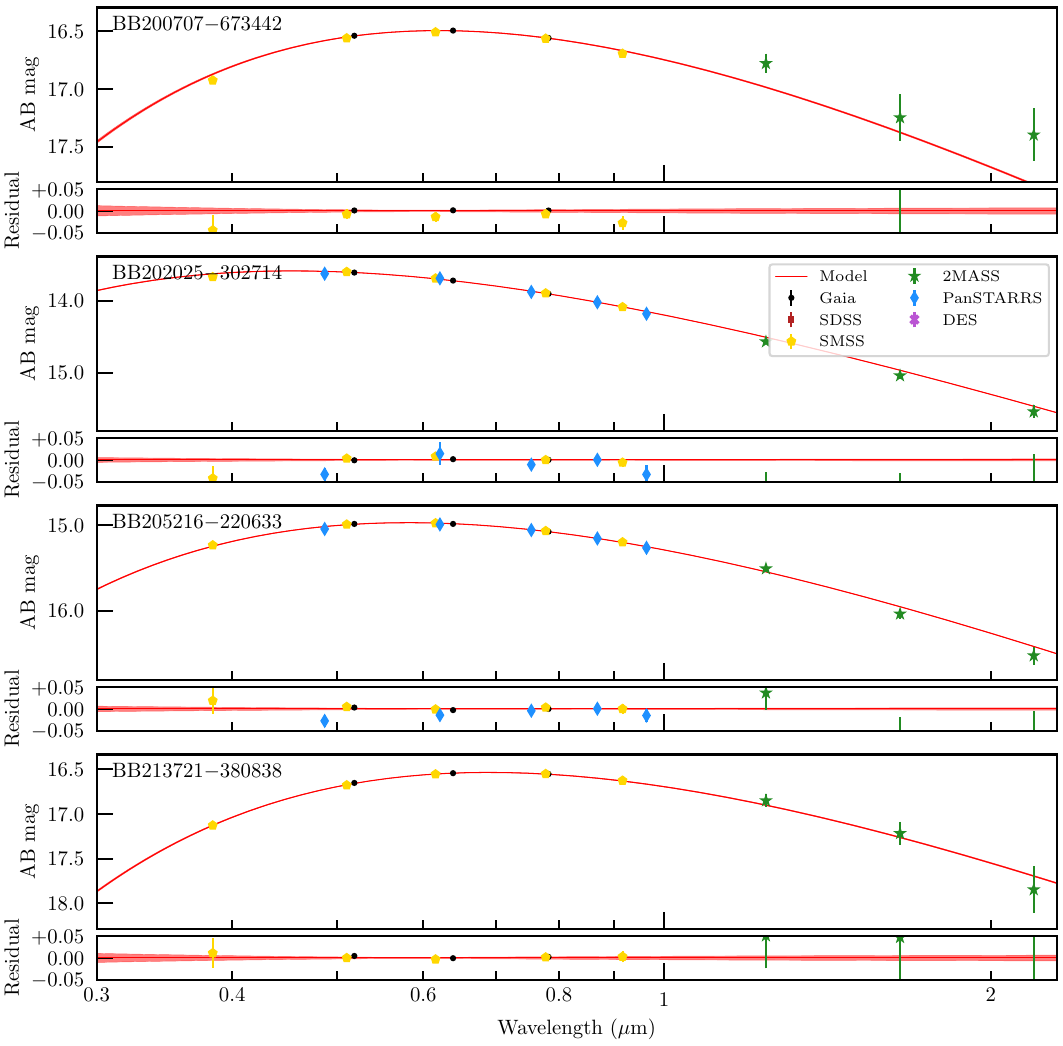}
    \caption{Same as Figure~\ref{fig:multipanelfits_1}, but showing the photometric fits to a different set of four featureless stars, as indicated in the top left corner of each panel.}
    \label{fig:multipanelfits_6}
\end{figure*}

In Figures~\ref{fig:multipanelfits_1}-\ref{fig:multipanelfits_6} we present the photometry and model blackbody fits to our featureless stars. We also present the fit residuals (data$-$model) below each main panel.

\bsp	
\label{lastpage}
\end{document}